\documentclass[12pt]{article}

\usepackage{array} 
\usepackage{epsfig}
\usepackage{amssymb}
\usepackage{amsmath}                          
\usepackage{graphics,graphpap}
\usepackage{graphicx}
\usepackage{epstopdf}

\renewcommand{\thefootnote}{\fnsymbol{footnote}}

\def\simge{\mathrel{%
   \rlap{\raise 0.511ex \hbox{$>$}}{\lower 0.511ex \hbox{$\sim$}}}}

\def\simle{\mathrel{
   \rlap{\raise 0.511ex \hbox{$<$}}{\lower 0.511ex \hbox{$\sim$}}}}

\newcommand{\aver}[1]{\langle #1\rangle}

\newcommand{\matel}[3]{\langle #1|#2|#3\rangle}
\newcommand{\state}[1]{|#1\rangle}
\newcommand{\astate}[1]{\langle #1|}

\newcommand{\dU}{{d_{\cal U}}}
\newcommand{\cU}{{\cal U}}

\begin{document}

\begin{titlepage}
\begin{flushright}\begin{tabular}{l}
IPPP/07/29\\
DCPT/07/58
\end{tabular}
\end{flushright}
\vskip1.5cm
\begin{center}
   {\Large \bf \boldmath Unparticles at heavy flavour scales: \\  CP violating phenomena}
    \vskip1.3cm {\sc
Roman Zwicky \footnote{Roman.Zwicky@durham.ac.uk}
  \vskip0.5cm
        {\em IPPP, Department of Physics, 
University of Durham, Durham DH1 3LE, UK}} \\
\vskip1.5cm 

{\em Version of \today}

\vskip2cm

{\large\bf Abstract:\\[8pt]} \parbox[t]{\textwidth}{
Coupling the scale invariant unparticle sector to flavour physics and
assuming that it remains scale invariant we investigate its 
consequences in heavy flavour physics. 
A  drastic feature of unparticle physics is an unusual phase leading 
to novel CP violating phenomena.
We consider the CP asymmetry in the leptonic decay $B^+ \to \tau^+ \nu$
and the hadronic decay  $B_d \to D^+D^-$, taking into account
constraints of branching ratios and time dependent CP asymmetries.
Generic plots are shown and it turns out that there exist parameters
for which the CP violation is maximal.
A prediction of a large CP asymmetry in $B_d \to D^+D^-$ is difficult to achieve in other models without contradicting the current data in other
channels.
The prediction of a CP asymmetry in leptonic decays, such as 
$B^+ \to \tau^+ \nu$, is novel. We identify the CP compensating
mode due to the unparticles and show explicitly that it exactly
cancels the CP asymmetry of $B^+ \to \tau^+ \nu$ as demanded
by CPT invariance.
Building up on earlier works we investigate the breaking of scale invaricance,
due to the coupling to the Higgs 
and the size of the effects in the weak sector resorting to a dimensional analysis.
An enhancement is observed on the grounds 
of the relevance of the
unparticle interaction operator as compared to the weak four-Fermi term.
}

\vfill

\end{center}
\end{titlepage}

\setcounter{footnote}{0}
\renewcommand{\thefootnote}{\arabic{footnote}}

\newpage

\section{Introduction}

The possibility of a non-trivial scale invariant sector, 
weakly coupled to the  Standard Model (SM), 
was advocated by Georgi in \cite{Georgi1}.
A scale invariant theory does not contain degrees of freedom
with isolated masses\footnote{In the absence of a mass scale the only possible particle
candidates seem to be  massless fields, but these have been shown to 
be free fields \cite{mass0free}.}, unlike field theories used in 
phenomenological particle physics.
Georgi called the degrees of freedom of such a theory
"unparticles".

A non-trivial scale invariant sector, i.e. non-vanishing coupling,
exhibits power-like scaling unlike the logarithmic scaling of QCD 
at the perturbative or trivial ultraviolet fixed point. 
The power-like scaling in Minkowski space seems to lead to curious phenomena.
For example, the phase space of an unparticle with scaling dimension $\dU$, which consists of the classical plus anomalous dimension, looks like a number of $\dU$  (possibly non-integer) massless
particles \cite{Georgi1}.  This could lead to interesting
signals of missing energy. In a second paper \cite{Georgi2} Georgi
has pointed out that the unparticle propagator 
has an unusual phase $e^{-i \dU \pi}/ \sin(\dU \pi)$ leading
to spectacular interference patterns. 

By parametrizing  a variety of interactions, unparticle phenomena
were investigated at  various energy scales and domains of 
particle physics such as
electroweak physics \cite{e+e-}, \cite{FRS}, \cite{WW},
collider physics \cite{DY}, \cite{collider_lot}, \cite{unparticle_resonance} (the latter investigates the (pseudo)resonance structure due to unparticles), DIS \cite{DIS}, \cite{nuN} $B$, $D$-physics \cite{Upheno},
 \cite{CPun} \cite{DDmix} \cite{BKnunu} \cite{CPun2}
 , light flavour physics \cite{PP}, \cite{mudecay}
$g_\mu\!-\!2$ \cite{Upheno}, \cite{gm2}
lepton flavour violation \cite{LFV1},\cite{LFV2}, invisible decays 
\cite{invisible}, cosmology \cite{cosmo} long-range interaction \cite{longrange} and gravity \cite{ungravity}; 
All studies are based on the assumption that the theory remains scale invariant until the respective energy domain.

Papers    addressing 
questions of interpretation and the range of scale invariance
have appeared.
In an illuminating paper by Stephanov \cite{Stephanov}
the continuous spectrum of
the unparticle fields is discretized  allowing for 
interpretation  in terms of the language of particle physics. 
The authors of reference \cite{FRS} address the question of
the range of scale invariance.
In the case where the unparticle couples to the Higgs  
vacuum expectation value (VEV) the latter will render
the theory non scale invariant.
This raises the question whether  unparticle effects 
are observable in low energy experiments. 
A follow-up paper has appeared \cite{BFRS}, where it
is observed that  higher dimensional operators can be
parametrically enhanced under certain conditions on the
scaling dimensions.
Moreover this paper contains
many physical applications and considers LEP-results to set bounds
on the effective suppression scale.

At first sight it seems rather difficult to pursue an analysis
in low energy physics. The unparticle effects are parametrized in 
terms of an effective field theory where no principle is (yet) known
to constrain the coefficients and a coupling to the Higgs VEV
would take the theory away from scale invariance.
On the other hand the novel phases could give rise
to such striking  phenomena that an 
investigation seems worthwhile.
Moreover we have adapted the analysis of Ref.~\cite{FRS}
to the weak sector and find that effects are possible if
the unparticle field couples weakly to the Higgs VEV.
The size of this coupling is not dictated by any principle 
and we may therefore regard its smallness as a
working assumption.

In the specific model or parametrization used, the unparticle will play the role of the 
$W$-boson or charged Higgs in flavour-changing decays.
It is well known that (time independent) CP violation manifests itself if 
there are at least two amplitudes with different relative strong (CP-even) and weak (CP-odd) phases.  
The phase of the unparticle propagator is CP-even and if we therefore allow for a different weak phase in the unparticle sector
the door is opened to novel CP violating phenomena. 
Decays with one dominant weak amplitude seem particularly suitable,
since they do not exhibit sizable CP violation. Moreover 
the unparticle should propagate at  large heavy flavour energies
because of the breaking of scale invariance. 
 
We analyze  leptonic decays of the type $B \to \tau \nu$, where the SM 
and  Beyond the Standard Models (BSM) do \emph{not} 
predict a CP asymmetry and $B_d \to D^+ D^-$, which is further motivated by the unexpectedly large CP asymmetry measured by the Belle collaboration \cite{BelleDD}.
We shall also investigate how large the impact of unparticles can 
be without conflicting with branching ratio and indirect CP asymmetry predictions.
 
The paper is organized as follows.
In section \ref{sec:scenario} the scenario of the model is described
including our parametrization of the effective Lagrangian and
some general notation for CP violation is introduced.
In section \ref{sec:decay} the leptonic decay $B \to \tau \nu$
and $B_d \to D^+D^-$ are investigated followed by a 
discussion of similar channels. 
In section  \ref{sec:cpt}  we verify a constraint on 
CP-violation from CPT-invariance; namely that  the partial sum of  particle
and antiparticle rates, with final states rescattering  into each other, 
are equal.
In section \ref{sec:dim}
we present the dimensional analysis of \cite{FRS} adpated
to a weak process. The paper ends with a summary and conclusions
in section \ref{sec:con}.
 
In this paper we shall adopt $\Lambda_\cU = 1\,{\rm TeV}$
as the scale of the IR fixed point. It is not difficult to rescale
the results to a different scale, in the relevant places
$(\Lambda_\cU/1\,{\rm TeV})$ will be shown explicitly
in the formulae.

\section{Scenario}
\label{sec:scenario}
According to \cite{Georgi1}, with slighly adapted notations from 
 \cite{FRS}, we shall imagine that  at a very high energy scale $M_\cU \gg 1\,{\rm TeV}$ the particle world is described by the
standard model fields and a self interacting ultraviolet sector.
These two sectors interact with each other via  heavy particles 
of mass $M_\cU$.
The ultraviolet sector is supposed to 
contain a non-trivial  infrared (IR) fixed point.
An example mentioned in \cite{Georgi1} is the Banks-Zaks \cite{BZ},
perturbative type, fixed-point. Other examples are gauge theories
with fermions in higher dimensional representations which exhibit
near conformality, also known as walking. The phase diagram for 
an arbitrary number of flavours and colours was given in \cite{HDR}
and preliminary lattice studies seem to confirm the theoretical
expectations \cite{HDRlattice}.
Below the scale $M_\cU$ the theory may be described by
non-renormalizable interactions
\begin{equation}
{\cal L}^{\rm eff} \sim 
\frac{1}{M_\cU^{d_{UV}+(d_{\rm SM}-4) }} O_{\rm SM} O_{UV} \quad ,
\end{equation}
analogous to the four-Fermi interactions connecting the
lepton and quark families within the SM.
The ultraviolet theory flows into the IR fixed point 
around some scale $\Lambda_\cU \sim 1\, {\rm TeV}$ which 
will lead to new degrees of freedom called unparticles described
by operators $O_{\rm IR} \equiv O_\cU$.
Below the  scale $\Lambda_\cU$  the theory may be described by an effective field theory in terms of the new degrees of freedom
\begin{equation}
\label{eq:effUn}
{\cal L}^{\rm eff} \sim \frac{\lambda}{\Lambda_\cU^{d_\cU+(d_{\rm SM}-4)}} O_{\rm SM} O_\cU 
\end{equation}
with coupling $\lambda = c_\cU (\Lambda_\cU / M_\cU)^{d_{UV}+(d_{\rm SM}-4)}$ and matching coefficient $c_\cU$. Since we do not 
have a concrete theory at hand $c_\cU$ will be a free parameter
to be constrained by experimental data.

To make use of Lagrangians of the form \eqref{eq:effUn} it will
prove sufficient to know the coupling of an unparticle degree of
freedom with given momentum $P$ to its field operator for calculating
decays into unparticles. Moreover via the optical theorem or 
dispersion relation it is possible to obtain the propagator and study interference effects, at the tree level, of unparticles and SM particles.
In this paper we are concentrating on the latter effect.

The propagator may be defined from its dispersion representation
\begin{eqnarray}
\label{eq:dispersive}
\Delta_\cU (P^2) \equiv  i \! \! \int_0^\infty d^4 x e^{i p\cdot x}
 \matel{0}{T O_\cU(x) O^\dagger_\cU(0)}{0} 
 = \int_0^\infty \frac{ds}{\pi} \frac{{\rm Im}[\Delta_\cU(s)]}{s-P^2-i0} + 
 {\rm s.t.}
\end{eqnarray}
where we have assumed that the unparticle state 
$|P\rangle$ satisfies $P^2 \geq 0$ and $P_0 >0$.
The abbreviation s.t. stands for possible subtraction terms
associated to a possible non convergence in the 
ultraviolet.
The imaginary part is given by
\begin{equation}
\label{eq:im}
{\rm Im}[\Delta_\cU(P^2)] = |\matel{0}{ O_\cU(0)}{P}|^2 = A_\dU (P^2)^{\dU-2}  \,.
\end{equation}
The $P^2$ dependence is solely determined by the scaling property of
the operator $O_\cU$. 
The factor $A_\dU$ is a priori an arbitrary normalization constant which has been chosen to be 
\begin{equation}
\label{eq:adu}
A_\dU = 16 \pi^{5/2} \frac{\Gamma(\dU+1/2)}{\Gamma(\dU-1)\Gamma(2 \dU)} 
\end{equation}
in reference \cite{Georgi1}. 
It is  the phase space volume of  $\dU$ massless 
particles. This choice was motivated by the fact that 
\eqref{eq:im} exhibits the same functional behaviour
as $\dU$ massless particles. This together with the fact that unparticles would (presumably) escape particle detectors has  led Georgi to point out \cite{Georgi1}:
"Unparticle stuff with scale dimension $\dU$  looks like a
non-integral number $\dU$ of invisible particles." 
With \eqref{eq:im} the dispersive integral \eqref{eq:dispersive} is elementary:
\begin{equation}
\label{eq:prop}
\Delta_\cU(P^2) = \frac{A_\dU}{2 \sin(\dU \pi)}\frac{1}{(-P^2-i0)^{2-\dU}} 
\stackrel{P^2>0}{\to} 
\frac{A_\dU}{2 \sin(\dU \pi)} \frac{e^{-i \dU \pi}}{(P^2)^{2-\dU}} \quad .
\end{equation}
We observe that the unusual phase, due to analtytic continuation to Minkowski space,
is due to the non-integral scaling dimension $\dU$.  
It has been shown in \cite{Georgi2} that the discontinuity of the 
propagator yields the imaginary part \eqref{eq:im}, which we have already implicitly assumed in the dispersion representation 
\eqref{eq:dispersive}. The phase is accompanied  by a term
$\sin(\dU \pi)$ in the denominator. In simple cases of direct CP asymmetries this term will cancel
and in more complicated cases it will partly cancel.
The cancelation of this term will play a crucial role when
we verify the the equality of partial rates of particles and antiparticles,
which is a consequence of CPT, in section \ref{sec:cpt}.
Plots of the function $A_\dU$ and $A_\dU/\sin(\dU \pi)$ can be found  in the appendix \ref{app:adu}.

A strong (CP-even) phase in a propagator appears 
spectacular and is not an element of  common models in particle physics. The phase factor
is due to dynamics in the unparticle or scale invariant sector.
The exactly solvable two-dimensional Thirring model, which
contains fermions with a current-current interaction term, is an example where the dynamical phase can be seen explicitly, c.f. appendix \ref{app:thirring}. The anomalous dimension is a function of the
coupling constant and assumes the free field value in the case where the 
coupling is taken to zero. We would like to add that in the Thirring model, due to the fermion selection rule, the anomalous phase is 
not so immediately observable.
It is also interesting to note that there is a connection between
the non-trivial phase and the causality structure.
The commutator of the unparticle field in the vacuum 
is, c.f. appendix \ref{app:comm} Eq.~\eqref{eq:comm},
\begin{equation}
\matel{0}{[O_\cU(x),O_\cU(0)]}{0} = -i \,{\rm sign}(x_0) \theta(x^2) (x^2)^{-\dU} f(\dU) \,,
\end{equation}
where $f(\dU)$ is a function explicitly given in \eqref{eq:comm},
The commutator vanishes for space-like $x^2 < 0$ separation
and obeys causality.
For generic $\dU$  the support is inside the light cone which seems
in agreement with the spectrum $P^2 \geq 0$. For an integer value, $\dU =  n$, the commutator behaves as $\sim \delta^{(n-1)}(x^2)$  and has support on the light cone only.
Note that in the latter case there is no CP-even phase in the unparticle
propagator.
  
We shall now discuss the possible values of $\dU$.
In the upper range $\dU = 2$ is singled out,
since we observe  that for $\dU \leq 2$ no subtraction terms  are needed. 
Moreover the singularity of $\dU$ approaching an integer value larger than $2$
has been interpreted in \cite{Georgi2} as the a $\dU$-particle cut which 
should not be attempted to be described by a single unparticle field.
In the lower range the value $\dU = 1$ seems special since 
it corresponds to the free massless field,
\begin{equation}
\label{eq:comm1}
\lim_{\dU \to 1} \Delta_\cU(P^2) = \frac{1}{P^2} \,.
\end{equation}
It is also observed that for $\dU \leq 1$ the dispersion 
integral does not converge in the infrared. This might be interpreted by
the fact that the field decreases even slower than the free massless field in coordinate space. It has been shown that for $\dU < 1$ the conformal group does not admit unitary representations \cite{Mack}.
Moreover in reference
\cite{Georgi1} it was noted that the decay into an unparticle 
has a non-integrable singularity in the decay rate for $\dU < 1$.
We shall therefore think of $\dU$ as being 
\begin{equation}
\label{eq:d12}
1 < \dU < 2 
\end{equation}
or parametrising with repsect to the free field limit,
$\dU = 1 + \gamma$ with anamalous  dimension
$0< \gamma < 1 $ at the non-trivial fixed point.

\subsection{Parametrization of the effective Lagrangian}

In this section we shall give our parametrisation of the
coupling of the unparticle sector to the SM.
We will investigate charged-flavour decays and therefore it is sufficient 
to give the couplings to that sector.
We couple a vectorial unparticle operator $O_\cU$ to a 
scalar and a pseudoscalar density.
The unparticle will therefore be a charged Lorentz-scalar and 
play the role of a charged Higgs  rather than a $W$-boson.
We parametrize the effective Lagrangian as 
follows\footnote{The channel  $B^+  \to \tau^+ \nu$ is mediated by a
$(P\!\times\! S) + (P\!\times \!P)$ structure whereas  
$B_d \to D^+ D^-$ decays via a $(S\! \times \!P)$ interaction. The vector  and axial  couplings are discussed in the text above and 
we do not consider tensor couplings since they 
do not couple to single scalar particles.}
\begin{eqnarray}
\label{eq:leff}
{\cal L}^{\rm eff} &=& \frac{\lambda_{S}^{ q' q}}{\Lambda_{\cal U}^{d_{\cal U}-1} } ( \bar q' q) \, O_{\cal U} \, +\,
 \frac{\lambda_{S}^{\nu l}}{\Lambda_{\cal U}^{d_{\cal U}-1} } (\bar \nu l) \, O_{\cal U} \,   \nonumber \\[0.1cm]
 &+& \frac{\lambda_{P}^{ q' q}}{\Lambda_{\cal U}^{d_{\cal U}-1} } (\bar q'\gamma_5 q) \, O_{\cal U} \, +\,
 \frac{\lambda_{P}^{\nu l}}{\Lambda_{\cal U}^{d_{\cal U}-1} } (\bar \nu
 \gamma_5l) \, O_{\cal U} \, +  \, {\rm h.c.} \,,
\end{eqnarray}
where $ q' = ( u,  c,  t)$, $q=(d,s,b)$,
$ \nu = (\nu_e,  \nu_\mu,  \nu_\tau)$ and   $l=(e,\mu,\tau)$ are summations over the families.
In the notation of   Eq.~\eqref{eq:effUn}, $d_{\rm SM} = 3$.
The weak (CP-odd) phases are parametrized as deviation from the 
phases of the CKM matrix $V_{\rm q'q}$ and analogously the
leptons as deviations from the PMNS matrix 
$U_{ \nu l}$.
\begin{equation}
\lambda_{S}^{q'q} = e^{i \phi^S_{q'q}}  | \lambda_{S}^{q'q} |   \qquad
\phi^S_{q'q} = {\rm arg}[{V_{q'q}}] + \delta \phi^S_{q'q} \quad .
\end{equation}

The Lagrangian is a non vectorial copy of the charged current sector in the SM.
This allows us to apply up to some level the same tools in the unparticle sector 
as in the SM. Note that the unparticle carries charge, unlike the Lagrangian
used in \cite{Georgi1}. 

We would like to stress that the Lagrangian in Eq.\eqref{eq:leff}, in the absence
of an explicit model realising the unparticle scenario, is not
dictated by any structure and  is therefore only an example.
Other Dirac structures with the same flavour transitions are possible.
The axial and vector structures, for example, can be coupled to a
transversal unparticle  $O^\mu_\cU$  ($\partial_\mu O_\cU^\mu=0)$ or to
a derivative coupling of a scalar unparticle  $\partial^\mu O_\cU$
\begin{eqnarray}
\label{eq:leffpart}
\delta {\cal L}^{\rm eff} &=& 
\frac{\lambda_{\partial(V,A)}^{q'  q}}{\Lambda_\cU^\dU }
 (\bar q' \gamma_\mu(\gamma_5) q) \, \partial^\mu O_{\cal U}  +  
 \dots
\end{eqnarray}
The former leads to a propagator transversal propagator  $\sim (-g_{\mu\nu} + p_\mu p_\nu/p^2)$ which vanishes by transversality when coupled to a pseudoscalar particle of momentum $p_\mu$, which is the case in the examples
considered in this paper.
The latter leads to an identical contribution as the scalar and pseudoscalar
contribution in the examples considered,  $\lambda_{S,P} \leftrightarrow   {\rm const} \cdot \lambda_{\partial (V,A)}$,
where $\lambda_{\partial (V,A)}$  is parametrically supressed by one
power of $\Lambda_\cU$ as compared to $\lambda_{S,P}$. We shall give 
the results in terms of $\lambda_{S,P}$ in the paper but also indicate explicitly 
how they change for a  $\lambda_{\partial (V,A)}$-coupling. 
Please note that although we have just stated that $\lambda_{S,P}$ and the 
$\lambda_{\partial (V,A)}$ are equivalent in the examples considered,
their role in model building, in regard to $SU(2)_{ L}$ for example,
might be rather different since the former couples
fermions of opposite chirality whereas the latter couples fermions of the same
chirality.


\subsection{General formulae for branching ratios and CP asymmetries}
\label{sec:genCP}

In this section we shall give the formulae for the branching 
ratio and CP asymmetries, used later on, in the case of
two amplitudes with different strong and weak phases.
This paragraph is completely general in principle, but we shall have 
in mind that one amplitude is due the SM and the other is due
to the unparticles.
For a decay $\bar B \to C D$ we parametrize
\begin{equation}
\bar
{\cal A}(\bar B \to X) = A_1 e^{i\delta_1} e^{i\phi_1} 
+ A_2 e^{i\delta_2}e^{i\phi_2} \quad ,
\end{equation}
where the $\delta_i$ denote the CP-even phases and 
$\phi_i$ the CP-odd phases. The branching ratio ${\cal B}$
and the CP averaged branching ratios are given by
\begin{equation}
{\cal B} =  {\cal B}^{ 0} \, f_{\Delta}
\qquad    {\cal \bar B} = {\cal B}^{0} \,   \bar f_{\Delta} \,,
\end{equation}
where 
\begin{eqnarray}
\label{eq:f}
{\cal B}^0 &=& \tau(\bar B) \frac{1}{16 \pi\, m_B^3} 
\lambda^{1/2}(m_B^2,m_{C}^2, m_{D}^2)
|A_1|^2 \qquad \Delta  = \frac{|A_2|^2}{|A_1|^2} \nonumber \\[0.1cm]
 f_{\Delta} &=&   (1+ 2 \Delta \cos(\phi_{12} + \delta_{12}) + \Delta^2) \nonumber  \\
\bar f_{\Delta} &=&  (1+ 2 \Delta \cos(\phi_{12} ) \cos(\delta_{12}) +\Delta^2)
\,,
\end{eqnarray}
and $\lambda(a,b,c) = a^2+b^2+c^2-2(ab+ac+bc)$, $\phi_{12} = \phi_1-\phi_2$ and
$\delta_{12} = \delta_1-\delta_2$.
In the case where  the transitions 
\begin{equation}
\label{eq:cond}
B \to   C D    \leftarrow \bar B
\end{equation}
are possible, the $B$-meson is neutral and the dynamical mixing of
$B_d$ and $\bar B_d$ leads to a time dependence in the CP asymmetry.
In the case where the coefficients $q$ and $p$, relating the flavour and
mass eigenstates of the neutral system, assume $|q/p| = 1$, 
the lifetime difference $\Delta \Gamma / \Delta M \ll 1$ the 
CP asymmetry assumes the following form
\begin{equation}
\label{eq:cpdef}
{\cal A}_{\rm CP}(B_d \to CD) \equiv 
\frac{\Gamma[\bar B \to  C  D] -\Gamma[ B \to  \bar C \bar D]}
{\Gamma[\bar B \to  C  D] + \Gamma[ B \to  \bar C \bar D]} = 
S_{CD} \sin(\Delta M t) - C_{CD} \cos(\Delta M t) \, .
\end{equation}
Both assumptions mentioned above are satisfied for the $B_d$ system. The sign convention is such that the $b \to c d$ decay 
rate enters with a plus sign, please note $\bar B \equiv \bar B_0 \equiv \bar B_d \sim (b\bar d)$ \cite{PDG}.
Writing $q/p = e^{-i \phi_{d}}$, where $\phi_d = 2 \beta$ is the mixing phase of 
the $B_d$ system,  $\lambda = q/p  ({\cal  A}/ \bar {\cal A})$,
then the (time independent)  CP asymmetry assumes the following form
\begin{equation}
\label{eq:C}
C = \frac{1-|\lambda|^2}{1+|\lambda|^2} = \frac{2 \Delta}{\bar f_\Delta} \sin(\delta_{12}) \sin(\phi_{12})
\quad .
\end{equation}
In the case where the system $C D$ is a CP eigenstate with eigenvalue $\xi_{CD}= \pm 1$ , 
which is a particular realization of \eqref{eq:cond},
the time dependent CP asymmetry assumes the following form 
\begin{equation}
\label{eq:S}
S = \xi_{CD} \frac{2 {\rm Im}(\lambda)}{1+|\lambda|^2} = \xi_{CD} \frac{-1}{\bar f_\Delta} ( \sin(\phi_d + 2 \phi_1)
+ 2 \Delta \cos(\delta_{12}) \sin( \phi_d + \phi_{12}) + \Delta^2 \sin(\phi_d + 2 \phi_2)) \,.
\end{equation}
For $B_d \to J/\Psi K_s$ $\xi_{J/\Psi K_s} = -1$, $\phi_1 \simeq 0$ and there is no sizable second amplitude in the SM 
and therefore $\Delta \simeq 0$ and  $\phi_{2} \to 0$ and
the gold plated formula $S_{J/\Psi K_s} = \sin(2 \beta)$ is recovered.

\section{A leptonic and a hadronic decay}
\label{sec:decay}

\subsection{$B^+ \to \tau^+ \nu $; scale invariant sector at $5\,{\rm GeV}$}

In the standard model charged pseudoscalars decaying to a lepton and a neutrino
are of particular interest because of their simple dependence on the
pseudoscalar decay constant and the CKM matrix element, see below. 

The novel feature when adding unparticles is a CP asymmetry. 
We will investigate how large this asymmetry can be, remaining consistent with the the branching ratio measurement

Below we will give  the decay amplitude for $B^+ \to \tau^+ \nu$ for the SM and
the unparticle contribution with effective Lagrangian as given in \eqref{eq:leff}.

\begin{figure}[h]
 \centerline{\includegraphics[width=5.2in]{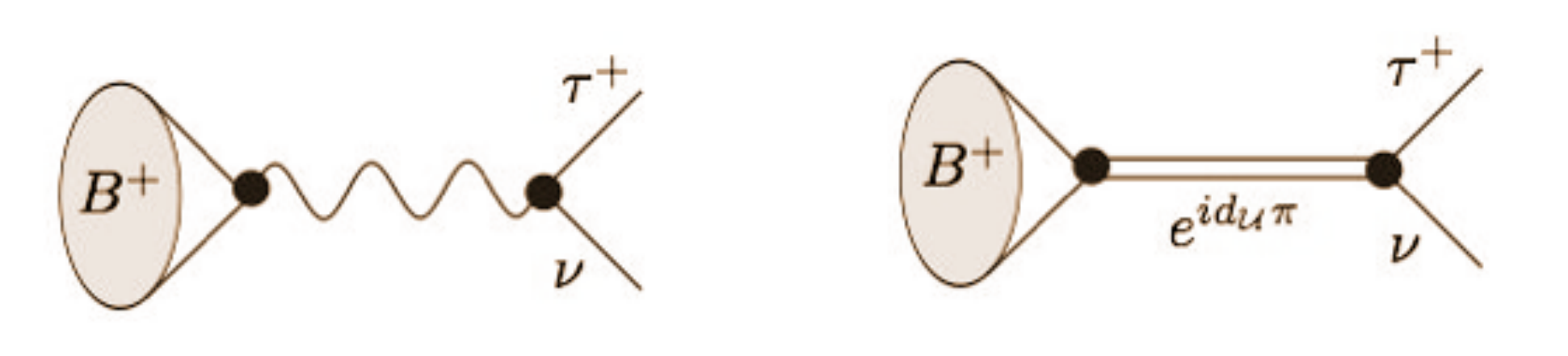}}
 \caption{\small (left) SM diagram for $B \to \tau \nu$ (right) unparticle
 diagram with CP odd phase $e^{i\dU\pi}$. The unparticle is denoted 
 by a double line.}
 \label{fig:btau_feyn}
 \end{figure}

The unparticle is propagating at the scale $m_B$ and we therefore assume
that the scale invariant sector  extends down to the $m_B$ scale.
The SM and unparticle graphs are shown in Fig.~\ref{fig:btau_feyn},
where we have indicated the phase. 

The additional unparticle amplitude leads to a slight complication. 
As a matter of fact in  experiment we do not observe the 
neutrino flavour but an inclusive measurement on the neutrino 
flavour is performed since the neutrinos are not detectable. 
In the case where there is only one amplitude, as  in the SM, unitarity of the PMNS matrix
hides this fact from the final formula. This is not the case for the unparticle amplitude
and we shall therefore derive formulae for 
$B^+ \to \tau^+ \nu \equiv \sum_l B^+ \to \tau^+ \nu_l$ via
$B^+ \to \tau^+ \nu_l$. The amplitude is the sum of two incoherent terms of opposite parity in the final state\footnote{The amplitude \eqref{eq:ampBtaunu} displays the famous helicity supression
in the SM due to its chiral structure which manifests itself in the fact that the amplitude 
is proportional to the lepton mass. For a pseudoscalar coupling, as in the charged Higgs
model, or the one used here, the helicity supression is relieved as can be inferred from Eq.~
\eqref{eq:deltatau}.
If we were to use a derivative coupling $\partial_\mu O$ to the axial vector 
\eqref{eq:leffpart} instead of a pseudoscalar coupling as in \eqref{eq:leff}
then the following substitution, $m_B^2/(m_b m_\tau) \to m_B^2/\Lambda_\cU^2$, in Eq.~\eqref{eq:deltatau} would   reproduce the result for the derivative coupling.}  
\begin{eqnarray}
\label{eq:ampBtaunu}
{\cal A}(B^+ \to \tau^+ \nu_l) &=& \frac{G_F}{\sqrt{2}} V_{\rm ub}^*
U_{\tau \nu_l} \, f_B   m_\tau \nonumber \cdot \\
& & \big( [\bar \nu \tau](1+  \Delta^S_{\tau\nu_l} e^{-i \dU \pi} e^{-i \phi^S_l})
+  [\bar \nu \gamma_5 \tau](-1+  \Delta^P_{\tau\nu_l} e^{-i \dU \pi} e^{-i \phi^P_l}
) \big) \,,
\end{eqnarray}
where $  \phi_l^D = \delta\phi^{P}_{\rm ub} - \delta\phi^{D}_{ \tau \nu_l}$ for $D = (S,P)$, $l = (e,\mu,\tau)$. The $B$-meson decay constant is defined as
$m_b\,\matel{0}{\bar b i\gamma_5 u}{B^+} = f_B m_B^2$ , where we  neglect isospin breaking effects.
The ratio of unparticle to SM amplitude is
\begin{eqnarray}
\label{eq:deltatau}
\Delta_{\tau\nu_l}^D &\equiv& 
 \frac{|\lambda_{D}^{\tau \nu_l }|}{|U_{\tau \nu_l}|} \tilde \Delta_{\tau \nu} \equiv
r_l^D  \tilde \Delta_{\tau \nu}  \nonumber  \,, \\[0.1cm]
 \tilde \Delta_{\tau \nu}  &= &   
\frac{|\lambda_{P}^{\rm ub}|}{|V_{\rm ub}|} \,
  \frac{A_\dU}{2 \sin(\dU \pi)} \frac{m_B^2}{m_b m_\tau}    \,
\frac{ (G_F/\sqrt{2})^{-1} }{m_B^2} \Big( \frac{m_B^2}{\Lambda_\cU^2}  \Big)^{\dU-1} \,.
\end{eqnarray}
We will now make a simplifying assumption in order to 
simplify the analyis.
We impose the left-handed chirality on the unparticle sector i.e. 
$\lambda_S^{\tau\nu_l} = - \lambda_P^{\tau\nu_l}$ and
($\Delta_{\tau \nu} \equiv \Delta_{\tau \nu}^{(S,P)}$,
$\delta\phi_{\tau \nu} \equiv \delta\phi_{\tau \nu}^{S}$, $r_l \equiv r_l^{(S,P)}$).
This means that
the amplitudes for opposite parity give the same result and this allows
us to combine the two amplitudes into one. 
The  branching fractions 
to a specific neutrino flavour  final state are
\begin{eqnarray}
& & {\cal B}(B^+ \to \tau^+ \nu_l)  =
{\cal B}_{\tau \nu}^{\rm SM} \, |U_{\tau \nu_l}|^2 f_{\Delta_{\tau\nu_l}} \qquad 
\bar {\cal B}(B^+ \to \tau^+ \nu_l)  =
{\cal B}_{\tau \nu_l}^{\rm SM} \, |U_{\tau \nu_l}|^2 \bar f_{\Delta_{\tau\nu_l}}  \,,
\end{eqnarray}
with $f$ and $\bar f$ as in \eqref{eq:f}, $\phi_{12} = -\phi_l$, $\delta_{12} = -\dU \pi$.
The familiar SM
branching fraction reads
\begin{equation}
\label{eq:SMBtaunu}
{\cal B}_{\tau \nu}^{\rm SM} = \tau(B^+) \, \frac{G_F^2}{8 \pi}\,  |V_{\rm ub}|^2 f_B^2 \, m_B m_\tau^2  (1-\frac{m_\tau^2}{m_B^2})^2
\end{equation}
and does not depend on the neutrino flavour. Please note that in the SM ${\cal B}_{\tau \nu}^{\rm SM} = \bar {\cal B}_{\tau \nu}^{\rm SM}$.
The experimentally tractable or neutrino inclusive branching fraction is 
\begin{eqnarray}
\label{eq:btau}
{\cal B}(B^+ \to \tau^+ \nu) &=&  \sum_l {\cal B}(B^+ \to \tau^+ \nu_l) 
\nonumber \\
& =& {\cal B}_{\tau \nu}^{\rm SM} \sum_l |U_{\tau \nu_l}|^2
 (1 + 2 r_l  \tilde \Delta_{\tau\nu} \cos(\phi_l + \dU \pi) + 
 (r_l  \tilde \Delta_{\tau\nu})^2) \nonumber \\
 & = &  {\cal B}_{\tau \nu}^{\rm SM} (1+  \sum_l |U_{\tau \nu_l}|^2
( 2 r_l  \tilde \Delta_{\tau\nu} \cos(\phi_l + \dU \pi) + 
 (r_l  \tilde \Delta_{\tau\nu})^2) \quad .
\end{eqnarray}
The formula could be further simplified if the $r_l$ were independent of $l$, which we shall assume shortly below. 
The CP averaged branching fraction is 
\begin{equation}
\label{eq:F}
\bar {\cal B}(B^+ \to \tau^+ \nu) \equiv  {\cal B}_{\tau \nu}^{\rm SM} {\cal F} =  {\cal B}_{\tau \nu}^{\rm SM}
(1+  \sum_l |U_{\tau \nu_l}|^2
( 2 r_l  \tilde \Delta_{\tau\nu} \cos(\phi_l ) \cos(\dU \pi) + 
 (r_l  \tilde \Delta_{\tau\nu})^2)) \, .
\end{equation}
The CP asymmetry assumes the following form 
\begin{equation}
\label{eq:acp}
{\cal A}_{\rm CP}(\tau \nu) \equiv
\frac{\Gamma(B^- \to \tau^- \bar \nu) -\Gamma(B^+ \to \tau^+ \nu)}
{\Gamma(B^- \to \tau^- \bar \nu) +\Gamma(B^+ \to \tau^+ \nu)} =
\frac{2 \tilde \Delta_{\tau\nu}}{\cal F}  \sin(\dU \pi)
\sum_l \sin(\phi_l) r_l |U_{\tau \nu_l}|^2 \,,
\end{equation}
where ${\cal F}$ is implicitly defined  in Eq.~\eqref{eq:F}.
Let us note that the CP violation encountered here is
proportional to $\sim{\rm Im} [V_{\rm ub}^* \lambda_P^{\rm ub} U_{\tau \nu} \lambda_{S}^{\tau \nu \, *}]$,
which is  hidden in the formula above,
and is the product of two quadratic reparametrization invariants.
The effect is entirely proportional to the sine of the phase difference
between the CKM (PMNS) and the unparticle flavour sector and can
therefore not occur in the SM.
In order to do a qualitative assessment we shall study the case where  there is no flavour dependent 
perturbation in the neutrino sector and therefore drop the label
$l$.
The formulae for the CP averaged branching ratio and the CP asymmetry then simplify to
\begin{eqnarray}
\label{eq:btau_analyze}
& & \bar {\cal B}(B^+ \to \tau^+ \nu)  \to {\cal B}^{\rm SM}_{\tau \nu}(1+  
 2  \Delta_{\tau\nu} \cos(\phi ) \cos(\dU \pi) + 
  \Delta_{\tau\nu}^2) \,  \stackrel{\phi = \pm \pi/2} {\to} \,  {\cal B}^{\rm SM}_{\tau \nu}(1+  \Delta_{\tau\nu}^2)
   \nonumber  \,,\\[0.1cm]
&  & {\cal A}_{\rm CP}(\tau \nu) \to \frac{2 \Delta_{\tau\nu}  \sin(\phi)  \sin(\dU \pi) }{1+  
 2  \Delta_{\tau\nu} \cos(\phi ) \cos(\dU \pi) + 
  \Delta_{\tau\nu}^2}  \,\stackrel{\phi = \pm  \pi/2} {\to} \,  \frac{\pm 2 |\Delta_{\tau\nu}|    |\sin(\dU \pi)| }{1+  
  \Delta_{\tau\nu}^2} \quad ,
\end{eqnarray}
where in the last step we have simplified the formulae further by setting the weak 
phase difference to  $90(270)^\circ\,$\footnote{N.B.  $\sin(\dU \pi) < 0$ for $1 <\dU < 2$ as assumed throughout this paper \eqref{eq:d12}.
This is the reason for the absolute values in the equation above.}.
N.B. in the notation used in Eq.~\eqref{eq:acp} 
${\cal A}_{\rm CP}(\tau \nu) = -C_{\tau\nu}$. 
This choice maximizes the CP violation for appropriate values for 
$\Delta_{\nu l}$.
Before we are able to constrain the CP violation with the rate we have to
give the theoretical and experimental results of the latter.

The following hadronic parameters,
$\tau^{B^+} = 1.643 \,{\rm ps}$, $f_B = (189 \pm 27) \,{\rm MeV}$ a lattice average from 
\cite{UTfit} and  $|V_{ub}| = 3.64(24) \cdot 10^{-3}$ from the fit to the angles of the
CKM triangle \cite{UTfit}, are used to estimate the SM branching fraction
\begin{equation}
\label{eq:btau_theory}
{\cal B}(B^+ \to \tau^+ \nu)^{\rm SM}_{\rm theory} = 83(40) \cdot 10^{-6} \, .
\end{equation}
We have doubled the uncertainty due to $|V_{\rm ub}|$.
This estimate has to be compared with the measurements at the $B$-factories

\begin{equation}
\label{eq:btau_exp}
\renewcommand{\arraystretch}{1.5}
\addtolength{\arraycolsep}{9pt}
\begin{array}{l | r | r | r | r |   }
  {\rm units}\, 10^{-6} & \bar{\cal B}(B^+ \to \tau^+ \nu )   \\
\hline  
{\rm BaBar}\cite{BaBarBtaunu}  (223{\rm M\, BB})    & 90(60)(10)    \\
{\rm Belle} \cite{BelleBtaunu}(449{\rm M\, BB})   &    179(53)(48) \\
\hline
{\rm HFAG} \cite{HFAG}        & 132(49)   \quad .
 \end{array}
\renewcommand{\arraystretch}{1}
\addtolength{\arraycolsep}{-9pt}
\end{equation}

\subsubsection*{Weak phase $\phi = 90(270)^\circ$, flavour independent perturbation neutrino sector}

\begin{figure}[h]
 \centerline{\includegraphics[width=3in]{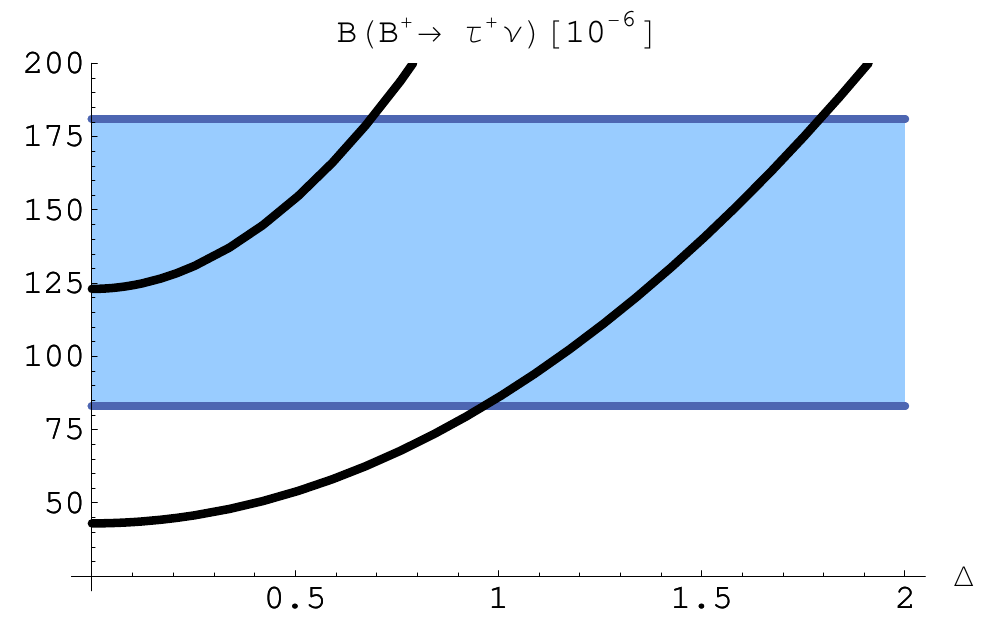},
 \includegraphics[width=3in]{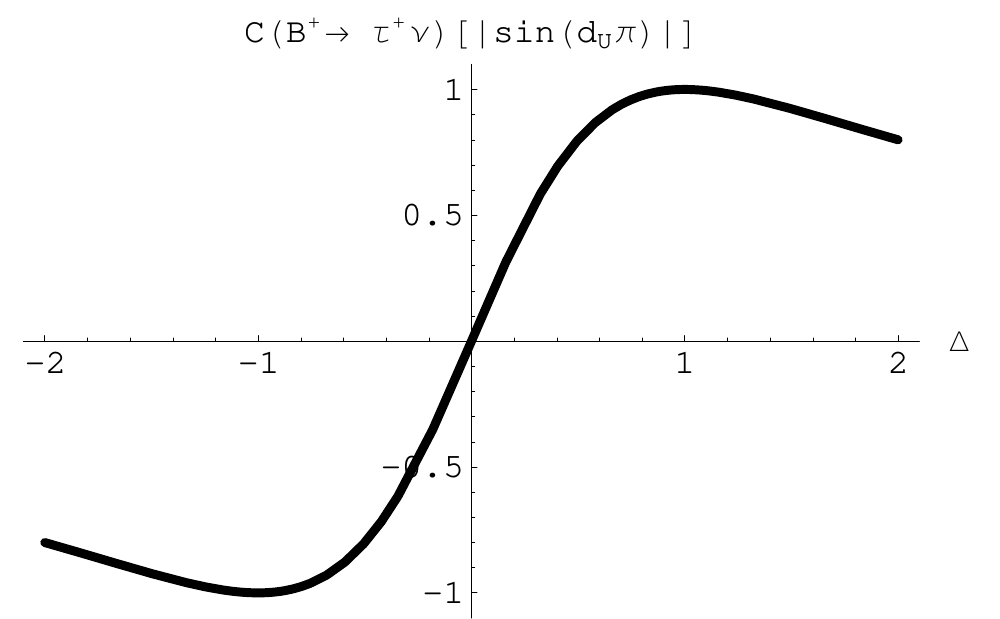}
 }
 \caption{\small   A weak phase difference $\phi = 90(270)^\circ$  is assumed here
 for $\Delta_{\nu\mu}$ positive(negative). 
 (left) Branching fraction \eqref{eq:btau_analyze} as a function of $\Delta_{\tau\nu}$.
 The black bands correspond to  the SM estimate
\eqref{eq:btau_theory} at  $\Delta_{\tau\nu}=0$. The blue band corresponds to the HFAG bounds
in Eq.~\eqref{eq:btau_exp}. (right)  The CP asymmetry as a function of  $\Delta_{\tau\nu}$ in units of $|\sin(\dU \pi)|$.
The scale $\Lambda_\cU = 1\,{\rm TeV}$ is chosen here.
N.B. in the notation used in Eq.~\eqref{eq:acp} 
${\cal A}_{\rm CP}(\tau \nu) = -C_{\tau\nu}$}
 \label{fig:btau} 
 \end{figure}

In Fig.~\ref{fig:btau} (left) the branching fraction 
\eqref{eq:btau_analyze} is plotted as
a function of $\Delta_{\tau\nu}$ with uncertainty taken from the SM estimate
\eqref{eq:btau_theory} at  $\Delta_{\tau\nu}=0$.
The blue band corresponds to the HFAG bounds
in Eq.~\eqref{eq:btau_exp}.
The CP asymmetry is plotted to the right of that figure.
The branching ratio does not set limits on the amount of CP violation, 
demanding the uncertainty bands to be tangent at worst $|\Delta_{\tau \nu}| < 1.8$.
Even in the case where the HFAG and theory uncertainty are halved, the 
value $|\Delta_{\tau \nu}| = 1$, at which the CP asymmetry is maximal, is 
still consistent.

\subsubsection*{Weak phase $\phi \neq 90(270)^\circ$, flavour independent perturbation neutrino sector}

In this subsection we shall repeat the analysis for a general 
weak phase difference and show two dimensional plots 
in the variables $(\phi,\dU)$ for different ratios of effective couplings.
The quantity $\Delta_{\tau\nu}$ \eqref{eq:deltatau}, used in the previous paragraph, depends on the 
ratio of effective coupling  and scaling dimension as follows
\begin{eqnarray*}
\label{eq:deltauexp}
\Delta_{\tau\nu}  &=&  \rho_{\tau \nu}
  \frac{A_\dU}{2 \sin(\dU \pi)} \frac{m_B^2}{m_b m_\tau}    \,
\frac{ (G_F/\sqrt{2})^{-1} }{m_B^2} \Big( \frac{m_B^2}{\Lambda_\cU^2}  \Big)^{\dU-1}   \\
&\simeq&  2300 \Big( 2.8\cdot10^{-5}  \frac{\Lambda_\cU}{1\,{\rm TeV}}  \Big)^{\dU-1} \frac{A_\dU}{\sin(\dU \pi)}  \rho_{\tau \nu} \,,
\end{eqnarray*}
where
\begin{equation}
\label{eq:rhotaunu}
 \rho_{\tau \nu} \equiv \frac{|\lambda_{P}^{\rm ub}\lambda_{(S,P)}^{\tau \nu} |}{|V_{\rm ub} U_{\tau \nu}|} \,.
\end{equation}
A plot relating $\Delta_{\tau\nu}$ and $\dU$ can be found in appendix
\ref{app:adu}, Fig.~\ref{fig:deltas}.
In Fig.~\ref{fig:btauBC} (right) the 
the CP asymmetry $C_{\rm \tau \nu}$  is plotted as a function of 
$(\phi,\dU)$ for $\rho_{\tau\nu} = (10^0,10^{-2},10^{-4})$.
The pattern is clearly regular and the 
condition for a large asymmetry is $|\Delta_{\tau\nu}| \sim 1$.
For smaller values of $\rho_{\tau\nu}$ the amount of possible CP violation is decreasing
because the condition mentioned above cannot be satisfied.
The constraint on the branching fraction, Fig.~\ref{fig:btauBC} (left), is defined by the following
acceptance function
 \begin{eqnarray}
 \label{eq:defi}
 A(\dU,\phi,\rho) &=& (1-r(\dU,\phi,\rho))\Theta(1-r(\dU,\phi,\rho)) \nonumber  \,,  \\[0.1cm]
r(\dU,\phi,\rho)  &=&  \frac{1}{\Delta B}\,  \big| {\cal B}^{\rm SM}_{\tau \nu}(1+ 
 2  \Delta_{\tau\nu} \cos(\phi ) \cos(\dU \pi) + 
  \Delta_{\tau\nu}^2)-  {\cal B}^{HFAG} \big|
 \end{eqnarray}
 for ${\cal B}^{\rm SM}_{\tau \nu} = 83\cdot 10^{-6}$, 
 $ {\cal B}^{\rm HFAG} =   132\cdot 10^{-6}$and
 for the quantity $\Delta B$ we add the uncertainty of
 the SM prediction and the HFAG value linearly  to
 $\Delta B \simeq  80\cdot 10^{-6}$. This function assumes values
 between $0$ and $1$, where $1$ signifies maximal agreement 
 and $0$ means that the point is excluded; or in other words we 
 consider predictions with a deviation larger than $\Delta B$ as excluded.
 
 For smaller values of $\rho_{\tau \nu}$ the linear term 
 for the branching ratio in Eq.~\eqref{eq:btau_analyze} 
 becomes dominant and a regular pattern in $\cos(\phi)$ 
 emerges. Note that since the predicted branching fraction 
 is lower than the central value from experiment,
 the weak angle  $\phi = 180^\circ$ is currently disfavoured 
 since it would lower the theory prediction even more.

\begin{figure}[!p]
 \centerline{\includegraphics[width=2.4in]{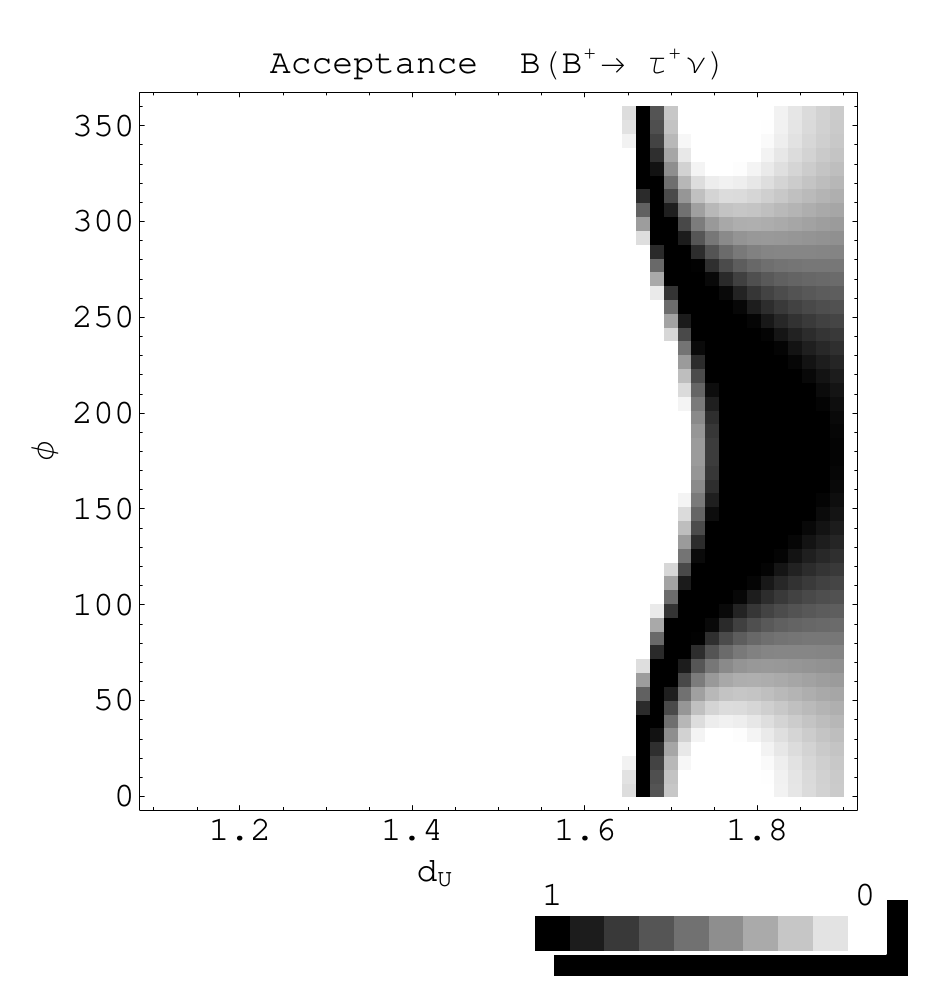}
 \includegraphics[width=2.4in]{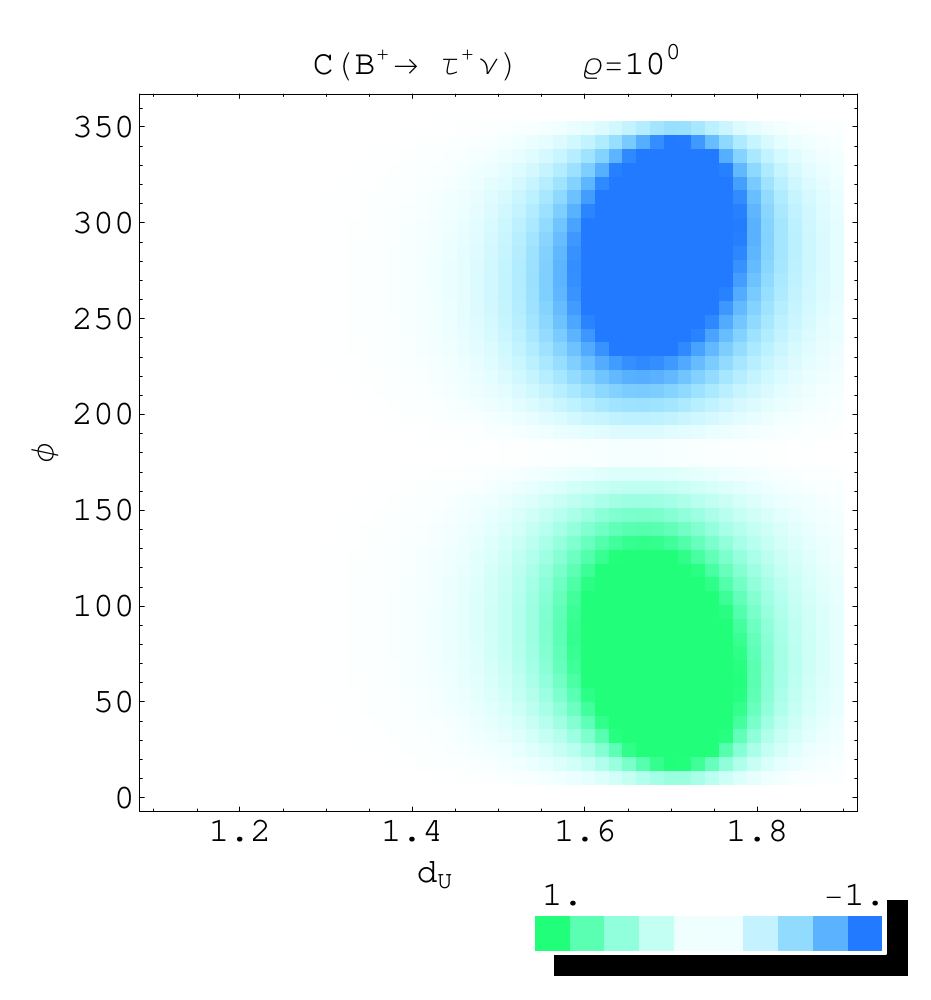}}
 \centerline{\includegraphics[width=2.4in]{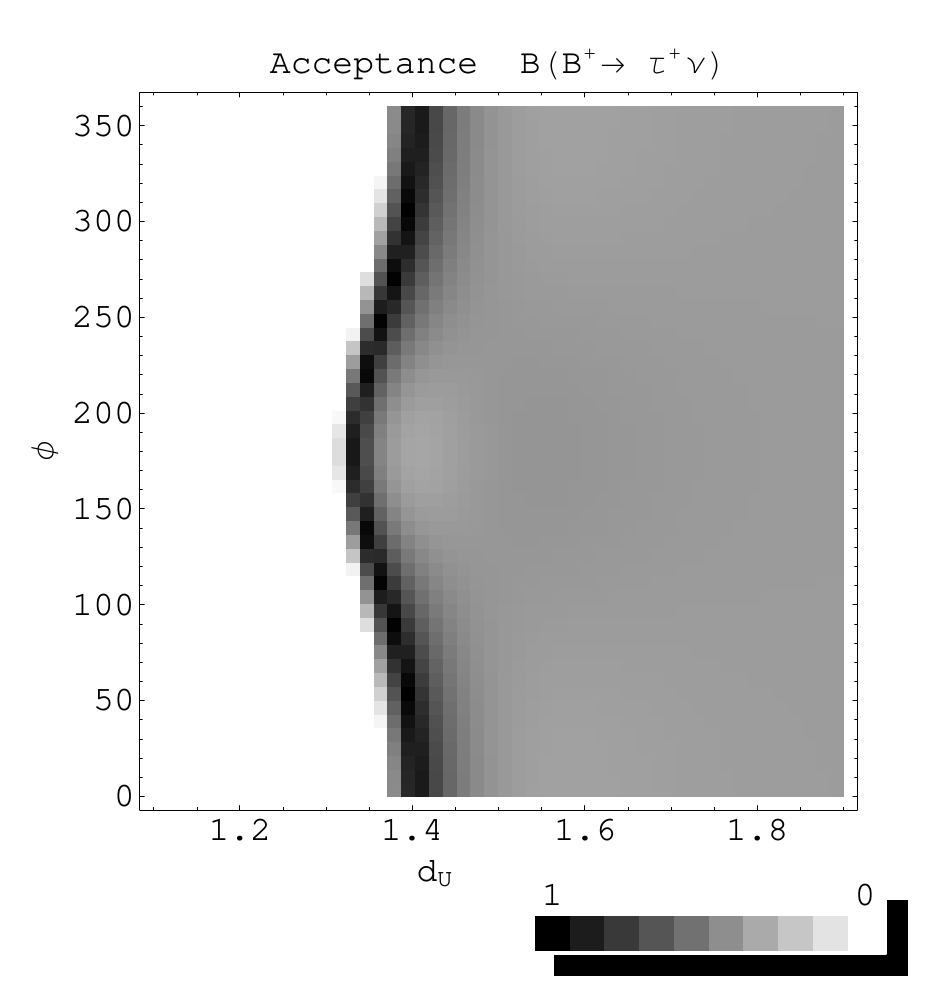}
 \includegraphics[width=2.4in]{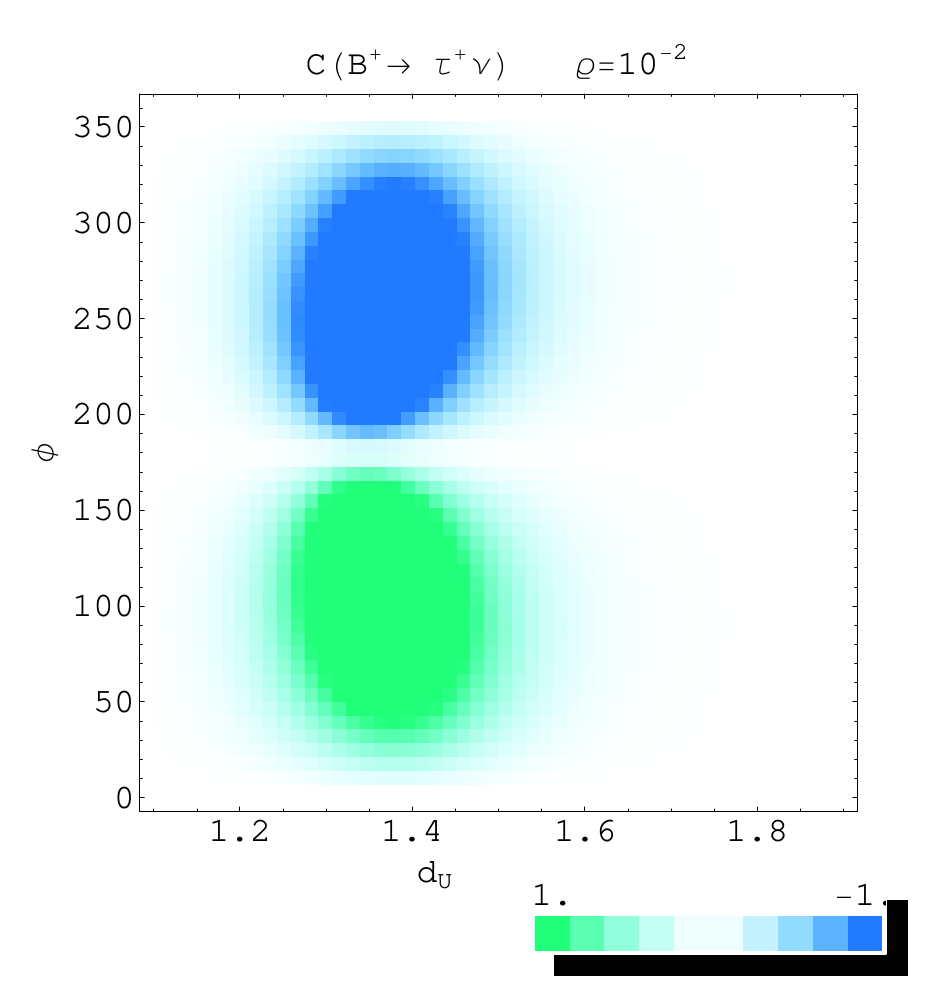}} 
 \centerline{\includegraphics[width=2.4in]{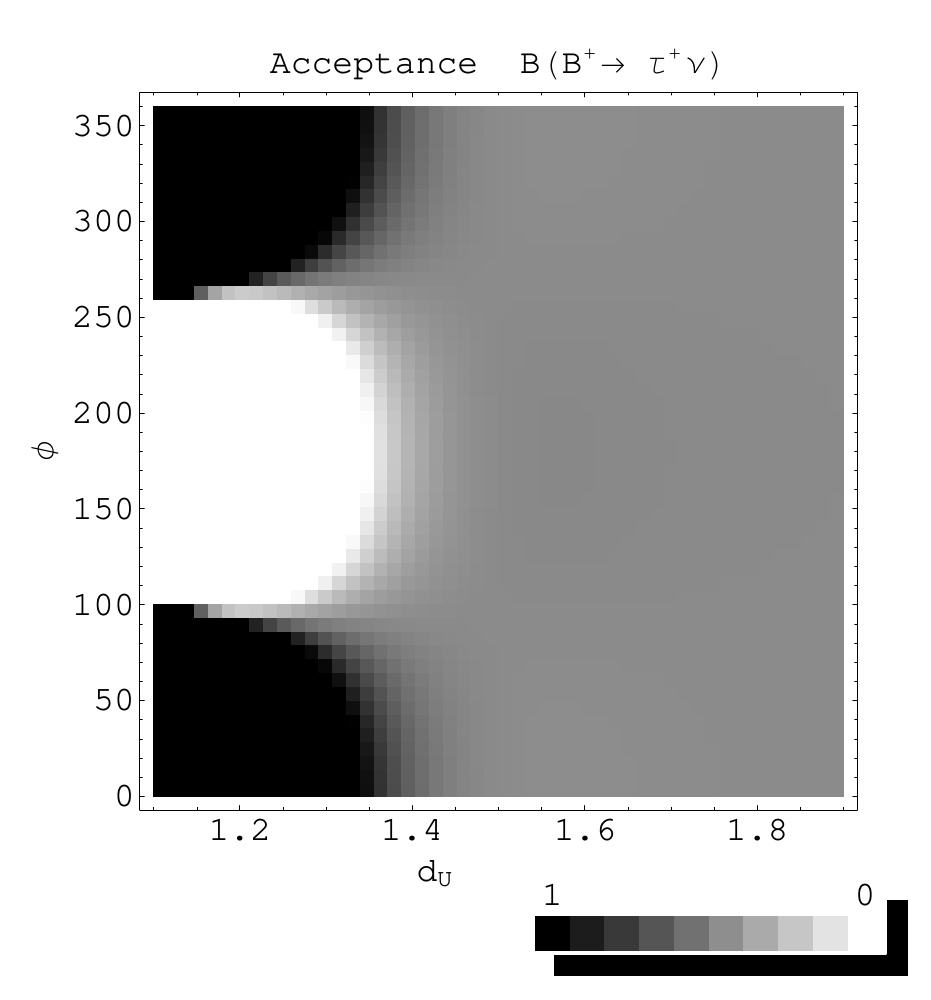}
 \includegraphics[width=2.4in]{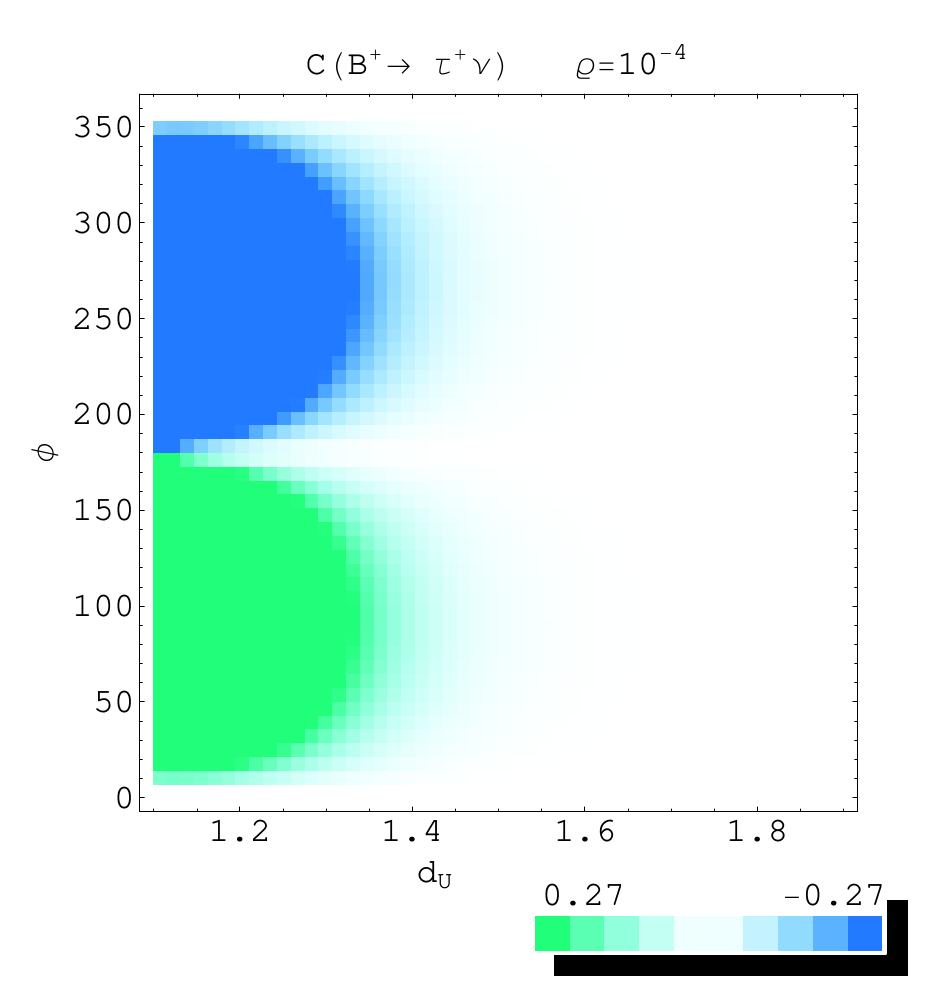}}
 \caption{\small A horizontal line of figures corresponds
 to different fractions of effective couplings
 $\rho = 10^{0,-2,-4}$ as defined in \eqref{eq:rhotaunu}.
 (left) Constraints on the $(\phi,\dU)$ parameter-space from the branching fraction. The values in the dark regions  are allowed whereas 
 white ones are excluded, c.f. Eq.~\ref{eq:defi} for a more details. (right) The CP asymmetry as a function of  $(\phi,\dU)$.
 The scale $\Lambda_\cU = 1\,{\rm TeV}$ is chosen here.}
 \label{fig:btauBC}
 \end{figure}
 
\subsection{Discussion and remarks on $B \to \mu \nu$, $D \to \mu \nu$, $B_s \to \mu^+\mu^-$ etc}
\label{sec:rembtaunu}
We have seen that applying the unparticle scenario to the 
leptonic decay $B \to \tau \nu$ leads to CP violation.
There is no experimental data available that gives 
both the negative and positive charged semileptonic decay
rates, i.e. quotes (bounds) on CP asymmetry in 
a semileptonic decay.

The current data on $B \to \tau \nu$ do not allow us to set 
bounds on the amount of possible CP violation. 
The amount of events at BaBar and Belle are of the order
$\sim 20$.
An improvement in theory, in particular on the $B$-meson
decay constant,  and the large statistics of a Super $B$-factory would 
of course improve the situation. Unfortunately the decay $B^+ \to \tau^+ \nu$
will not be possible or competitve at LHCb because of the neutrino final state and
the intricacies in the $\tau$ detection, whether  $D(D_s) \to (\tau,\mu) \nu$
decays are possible at LHCb is currently under investigation.

We shall comment on other leptonic modes. They are all 
described by the same formula \eqref{eq:SMBtaunu}
for $B \to \tau \nu$ with obvious substitutions for
$V_{\rm ub}$, $f_B$, $m_B$ and $m_\tau$.
We may also consider the $D$-decays assuming that the scale invariant sector  extends  to $\sim 2\,{\rm GeV}$.
The decay $D^+ \to \mu^+ \nu$ is measured by CLEO \cite{CLEO} the $\sim 50$ events 
lead to a thirteen percent accuracy.
The decay constant $f_D^+ = 220(20)\, {\rm MeV}$ is taken as an average
value of theory determinations from the table in \cite{CLEO}
and $|V_{\rm cd}| = 0.227$ \cite{PDG}. The Cabibbo allowed decays 
$D_s^+ \to \mu^+ \nu$ are measured as well \cite{PDG}, although with less precission. The decay constant $f_{D_s}^+  =  264(36)$ is obtained from an
average of $f_{D_s}^+/f_D^+  = 1.20(5)$ of the table in \cite{CLEO} and
$|V_{\rm cs}| = 0.957(17)(93)$ \cite{PDG}.
 A  summary  of the experimental \cite{PDG} and  theory
predictions is:
\begin{equation*}
\renewcommand{\arraystretch}{1.5}
\addtolength{\arraycolsep}{9pt}
\begin{array}{l | r | r | r  }
                &   B \to \tau \nu &  B \to \mu \nu &  B \to e \nu  \\ 
\hline  
{\rm Experiment} &  132(49) \cdot10^{-6}  & < 17 \cdot10^{-7} & < 9.8\cdot 10^{-6}  \\
{\rm Theory} & 83(50\%) \cdot 10^{-6}   &  3.7(50\%) \cdot10^{-7} & 8.4(50\%) \cdot 10^{-12}             \\
\hline
               & D \to \tau \nu & D \to \mu \nu & D \to e \nu  \\ 
               \hline
 {\rm Experiment}               &   <  2.1\cdot10^{-3}   &   4.4(7)\cdot 10^{-4} &  <  2.4\cdot 10^{-5} \\  
 {\rm Theory}               &   1.1(20\%) \cdot 10^{-3}  & 4.3(20\%) \cdot10^{-4} &   1.0(20\%) \cdot 10^{-10} \\
 \hline
               & D_s \to \tau \nu & D_s \to \mu \nu & D_s \to e \nu  \\ 
               \hline
 {\rm Experiment}               &   6.4(15)\cdot10^{-2}   &   6.3(18)\cdot 10^{-3} &  {\rm not\,\,available} \\  
 {\rm Theory}               &   5.5(30\%) \cdot 10^{-2}  & 5.7(30\%) \cdot10^{-3} &   1.3(30\%) \cdot 10^{-7}
   \,,
 \end{array}
\renewcommand{\arraystretch}{1}
\addtolength{\arraycolsep}{-9pt}
\end{equation*}
The $B$ decays are predicted to $50\%$ due to uncertainties in $f_B$ and $|V_{\rm ub}|$,
whereas the $D(D_s)$ decays have a lower uncertainty $20(30)\%$ due to $f_D(f_{D_s})$. 
The helicity  supression in the SM is apparent from the table.

Repeating the analysis for $D^+ \to \mu^+ \nu$, as shown in Fig.~\ref{fig:btau}, 
we obtain that $|\Delta_{D \to \mu \nu} | < 0.65$ which still allows 
for a rather large CP asymmerty, $|C_ {D \to \mu \nu} | < 0.9$.

The prediction of these modes in the SM is solid and a significant deviation
would be a clear hint for new physics. In particular one expects larger
rates in models where the helicity suppression is relieved.
An example is  the charged Higgs or the effective Lagrangian used in 
this paper. 
The charged Higgs does not predict a significant 
CP asymmetry whereas in unparticle models it is possible and 
therefore a CP asymmetry could be used to discriminate between the models.

We would also like to mention the decay $K^+ \to \mu^+ \nu$,
the KLOE collaboration reports $\sim 860$ events and a branching 
ratio ${\cal B}(K^+ \to \mu^+ \nu(\gamma)) =0.6366(9)(15)$ \cite{KLOE}.
On the one hand it seems unreasonable that the scale invariant sector 
 could extend to $\sim 500\,{\rm MeV}$ but on the other  
this channel has the largest statistics. If we assume that theory predicts the
rate to $5\%(10\%)$ this would roughly bound $|\Delta_{K \to \mu\nu}|< 20(30)$
and the CP asymmetry to $|C_{K \to \mu\nu}| < 0.4(0.55)$.

Finally a comment about $B_{(d,s)} \to \mu^+\mu^-$. 
This  channel is rare since it is a flavour-changing neutral decay further
 suppressed by the  coupling of the $Z$ and the helicity of final states, ${\cal B}(B_{(d,s)} \to \mu^+\mu^-)^{\rm SM} \sim 10^{-10}(10^{-8})$.
The branching ratio is not yet measured, the bounds are about one and half order of magnitude away
from the SM prediction. An analysis along the lines of $B \to \tau \nu$ does not make sense since there
are no direct constraints in that channel. A possibility would be to combine it with 
constraints from $\Delta M_{(d,s)}$, which are measured, as  advocated in reference \cite{CPun}.


\subsection{$B_d \to D^+ D^-$; scale invariant sector at $2\,{\rm GeV}$}

The decay $B_d \to D^+D^-$ corresponds to a $b \to \bar c c d$ transition at the quark level and 
is colour allowed.
It has the same  quark level transition as $B_d \to J/\Psi \pi_0$ but two complications arise
as compared to the latter.
First, since it is colour  
allowed it receives
sizable contributions from a gluonic penguin\cite{BSS}
and second the final states 
combine into a sum of isospin $I=0$ and $I=1$ waves which have in general different final state
interaction phases. 
Ultimately  we will neglect the penguins in our analysis, to be discussed below.
Our motivation to investigate the $B_d \to D^+D^-$ is 
driven by the measurement of a large CP asymmerty by the 
Belle collaboration \cite{BelleDD}\footnote{I am grateful to Christopher Smith for drawing my attention to this measurement.}. 
The SM expectation is  $C_{D^+D^-}^{\rm SM} \simeq -0.05$.

\begin{equation}
\label{eq:DDCP}
\renewcommand{\arraystretch}{1.5}
\addtolength{\arraycolsep}{9pt}
\begin{array}{l | r |  r  }
  & C_{D^+D^-} & S_{D^+D^-}   \\
\hline  
{\rm BaBar}\cite{BaBarDD}  (364{\rm M\, BB})    &  0.11(22)(07)  & -0.54(34)(06)\\
{\rm Belle} \cite{BelleDD}(535{\rm M\, BB})     &     -0.91(23)(06)& -1.13(37)(09) \\
\hline
{\rm HFAG}            &     -0.37(17) &                     -0.75(26)  
 \end{array}
\renewcommand{\arraystretch}{1}
\addtolength{\arraycolsep}{-9pt}
\end{equation}
It has to be said that the Belle result is 
somewhat moderated by a significantly lower value from BaBar \cite{BaBarDD} 
with opposite sign.
Note that the cental 
values from Belle also violate the general bound $C^2 + S^2 \leq 1$.

It shall be our goal to see  how large a CP asymmetry $C_{D^+D^-}$ 
the unparticles scenario can generate and still be consistent
with the branching fraction and the time dependent CP asymmetry.

\begin{figure}[!h]
 \centerline{\includegraphics[width=5in]{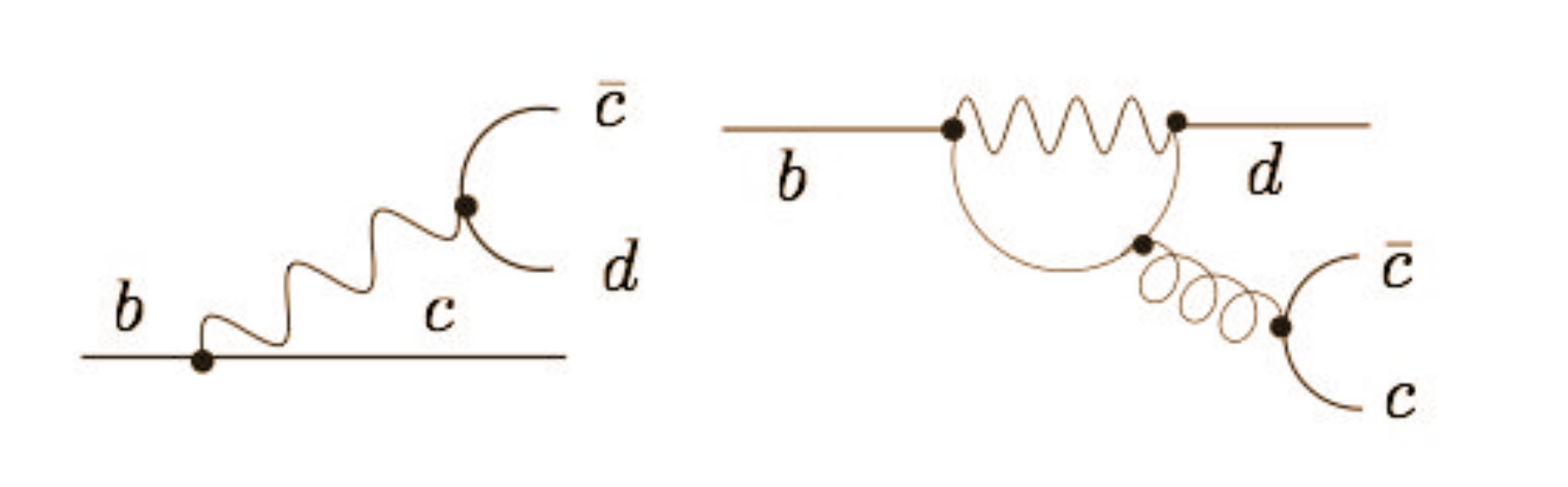}
 }
 \caption{\small $b \to d \bar c c$  (left) tree diagram,  (right) penguin diagram  }
 \label{fig:treeANDpenguin}.
 \end{figure}

In our analysis the  unparticle will replace the $W$ in the tree level amplitude
in, c.f. Fig.~\ref{fig:treeANDpenguin} (left). We therefore assume that the scale invariant sector 
extends to the $D$-meson scale  $\sim 2 \,{\rm GeV}$.

We shall first reconsider the situation in the SM before we move on to the unparticles.
Writing the amplitude as the sum of the tree
and penguin topology
\begin{equation}
{\cal A}(B_d \to D^+D^-) = {\cal A}_T + {\cal A}_P = {\cal A}_T(1-  e^{i\delta_{\rm PT}}e^{i \gamma} r_{\rm PT}) \,,
\end{equation}
the ratio of penguin to tree amplitude $r_{\rm PT}$ can then be estimated by the
Bander-Silverman-Soni mechanism \cite{BSS}, c.f. \cite{Xing} or \cite{Fleischer} for an updated analysis, 
\begin{equation}
\label{eq:resi}
\Delta_{\rm PT} \simeq 0.08 \qquad \delta_{\rm PT} \simeq 205^\circ \quad .
\end{equation}
This allows us to obtain the asymmetries from \eqref{eq:C} and \eqref{eq:S},
\begin{equation}
C_{D^+D^-}^{\rm SM} \simeq -0.05 \qquad S_{D^+D^-}^{\rm SM} \simeq -0.78 \quad .
\end{equation}
Comparing with the experimental results \eqref{eq:DDCP} we infer that the SM
is in good agreement with the time dependent CP asymmetry $S_{D^+D^-}$. 
The direct CP asymmetry $C_{D^+D^-} \simeq 0.05$ is 
about two standard deviations lower than the HFAG value $0.37(17)$.
In view of the non consistency of the two measurements it is certainly
wise to wait for updates from the $B$-factories.
We will in the following neglect the penguin contribution in regard to its
moderate size \eqref{eq:resi} in the SM. We will also neglect the 
"unparticle penguin". The ratio of the unparticle penguin amplitude
to the unparticle amplitude is expected to be of the same size as
in the SM, unless the up-type transition is enhanced by the
effective couplings. We are therefore implicitly assuming that
$|\lambda_{(S,P)}^{\rm ub} \lambda_{(S,P)}^{\rm ud}| \simle |\lambda_{(S,P)}^{\rm cb} \lambda_{(S,P)}^{\rm cd}|$.

We will describe the amplitude  $B_d \to D^+D^-$ within the naive factorization approximation.
Naive factorization describes colour allowed modes (topology as in 
Fig.~\ref{fig:treeANDpenguin} to the left) like $B \to \pi^+\pi^+$ and 
$B_d \to D^+ \pi^-$ with at least one fast or light meson with an accuracy of around
$10-20\%$ level.  For $B_d \to D^+D^-$, factorization in general and naive factorization
are not expected to hold. The overlap of the emitted $D^+$-meson with the $B_d \to D^-$
transition is expected to be relatively large. 
However it is empirically observed that naive factorization still works reasonably well.
We shall account for final state interactions, not included in naive factorisation,
 by an isospin analysis which is presented in the appendix \ref{app:iso}.
The effect is that the amplitude  receives a contribution $\cos((\delta_1-\delta_0)/2) \simeq \pm 0.63(15)$, c.f. 
\eqref{eq:iso_final}.
In fact the sign is not determined but since it enters in the square in the observables
it is of no concern here. 
The amplitude for $B_d \to D^+D^-$ in the SM is
\begin{eqnarray}
\label{eq:BDDamp}
{\cal A}(B_d \to D^+ D^-)  &=& \frac{G_F}{\sqrt{2}} V_{\rm cb}^*V_{\rm cd}
a_1 f_D \big( (m_B^2-m_D^2)  f_+^{BD}(m_D^2) + 
m_D^2 f^{BD}_-(m_D^2)\big) \nonumber \\
& \times  & \cos\big( (\delta_1-\delta_0)/2 \big) e^{i (\delta_1-\delta_0)/2}    \equiv   {\cal A}^{\rm SM}_{\rm DD} 
\,  ,
\end{eqnarray}
where $a_1 = C_2 + C_1/3 \simeq 1$ is the colour allowed combination of tree level Wilson coefficients and the $D$-meson
decay constant is defined as $m_c\,\matel{0}{\bar c i\gamma_5 d}{D^-} = f_D m_D^2$, where we neglect effects due to isospin breaking.
The $B \to D$ form factor can be parametrized by use of Lorentz covariance as 
\begin{equation}
\label{eq:FF}
\matel{D}{\bar b \gamma_\mu c }{B} = f^{BD}_+(q^2)(p_B+p_D)_\mu
+ f^{BD}_-(q^2) q_\mu \,,
\end{equation}
with momentum transfer $q = p_B-p_D$. The form factors are related to the famous
Isgur-Wise function  $f^{BD}_+(q^2) = \sqrt{\frac{m_B+m_D}{4 m_B m_D}} \xi(w)$,
$f^{BD}_-(q^2) = -\sqrt{\frac{m_B-m_D}{4 m_B m_D}} \xi(w)$ in the heavy quark limit.
Here  $w = v\cdot v'= (m_B^2+m_D^2 - q^2)/(2 m_B m_D)$. Whereas the normalization of the Isgur-Wise function $\xi(1)=1$
follows from charge normalziation in the heavy quark limit the values around
maximum recoil are much less known. We shall take the value $f^{BD}_+(0) =0.54$ from
\cite{Roma} and scale it up to $q^2 = m_D^2$ by use of a single pole model \cite{NeubertStech},
$\xi(w) \sim \sqrt{2/(w+1)} (w_{\rm max} - w(m_{B_c^*}^2)/(w-w(m_{B_c^*}^2)$.
The $B^*_c$-meson has the correct quantum numbers $J^P=1^+$ and its mass is the same in the heavy quark limit as  $ m_{B_c} = 6.29\,{\rm GeV}$ \cite{PDG} .
We obtain  $f_+(m_D^2) \simeq 0.7$. With $f_D = 220 \,{\rm MeV}$.
we get 
\begin{equation}
\label{eq:BDD_theory}
\bar {\cal B}(B_d \to D^+ D^-)^{\rm SM}_{\rm theory} = 1.7(10) \cdot 10^{-4}
\end{equation}
as a theory estimate, where the bulk of the uncertainty quoted is due to the isospin final state 
interaction phases \eqref{eq:iso_final}.
This estimate has to be compared to the experimental value \cite{PDG}
\begin{equation}
\label{eq:BDD_exp}
\bar {\cal B}(B_d \to D^+ D^-)_{\rm PDG} = 1.9(6) \cdot 10^{-4} \quad .
\end{equation}
The agreement seems accidentally good in regard to 
the approximations made.

As in the previous section we parametrize the amplitude
\begin{equation}
{\cal A}(B_d \to D^+ D^-) \equiv  {\cal A}^{\rm SM}_{\rm DD}     \big(1 + \Delta_{DD} e^{-i\phi_\cU} e^{-i \phi }  \big)
\end{equation}
with ${\cal A}^{\rm SM}_{\rm DD}$ as given in \eqref{eq:BDDamp} and relative weak phase
$\phi \equiv \delta \phi_{\rm cb} - \delta \phi_{\rm cd}$.  The ratio of SM to unparticle amplitude is\footnote{A derivative coupling $\partial_\mu O_\cU$ 
to  the vector Dirac-structure \eqref{eq:leffpart}
 instead of a scalar coupling as in \eqref{eq:leff} would lead to 
 a change of 
$m_D^2/(m_c (m_b-m_c)) \to m_D^2/\Lambda_\cU^2$ in 
Eq.~\eqref{eq:deltaDD}.}
\begin{eqnarray}
\label{eq:deltaDD}
\Delta_{\rm DD} &=&  \frac{|\lambda_{S}^{\rm cb}\lambda_{P}^{\rm cd} |}{|V_{\rm cb} U_{\rm cd}|} \frac{1}{a_1} 
 \frac{A_\dU}{2 \sin(\dU \pi)} \, \frac{m_D^2}{m_c(m_b-m_c)} \,
\frac{(G_F/\sqrt{2})^{-1}}{m_D^2} \Big( \frac{m_D^2}{\Lambda_\cU^2} \Big)^{\dU-1}   \,.
\end{eqnarray}
Note that, unlike for $B \to \tau \nu$, the negative parity of the $D$-meson selects only the $\lambda_P^{cd}$ coupling in the final vertex. 
The observables are obtained from Eq.~\eqref{eq:C} and \eqref{eq:S}  with 
$\xi_{D^+D^-} = 1$, $\phi_d = 2\beta$, $\phi_1= 0$, $\phi_2 = -\phi$ and $\delta_{12} = d_U \pi$ :
\begin{alignat}{2}
\label{eq:BDD_analyze}
& {\cal B}_{\rm DD} \,&=&\, {\cal B}_{\rm DD}^{\rm SM} \, f_{\Delta_{\rm DD}}
\qquad    {\cal \bar B}_{\rm DD} = {\cal B}_{\rm DD}^{\rm SM} \,   \bar f_{\Delta_{\rm DD}}
\nonumber \,,  \\[0.1cm]
& C_{\rm DD} \,&=&\, 
\frac{2 \Delta_{\rm DD}}{\bar f_{\Delta_{\rm DD}} } \sin[\phi] \sin[\dU \pi] \nonumber \,, \\[0.1cm]
& S_{\rm DD} \,&=&\, 
\frac{-1}{\bar f_{\Delta_{\rm DD}} }
 (\sin[2 \beta]  +   2\Delta_{\rm DD} \cos[\dU \pi] \sin[2 \beta - \phi]
 + \Delta_{\rm DD}^2  \sin[2 \beta - 2\phi])
\end{alignat}
and
\begin{equation}
{\cal B}_{\rm DD}^{\rm SM} =   \tau(B_d) \,\frac{G_F^2}{32 \pi m_B }  a_1 ^2 f_D^2 
 \big( (m_B^2-m_D^2)  f_+^{BD}(m_D^2) + 
m_D^2 f^{BD}_-(m_D^2)\big) ^2
 |V_{\rm cb}^*V_{\rm cd}|^2  \, .
\end{equation}

\subsubsection*{Weak phase $\phi = 90(270)^\circ$}

\begin{figure}[h]
 \centerline{\includegraphics[width=2.2in]{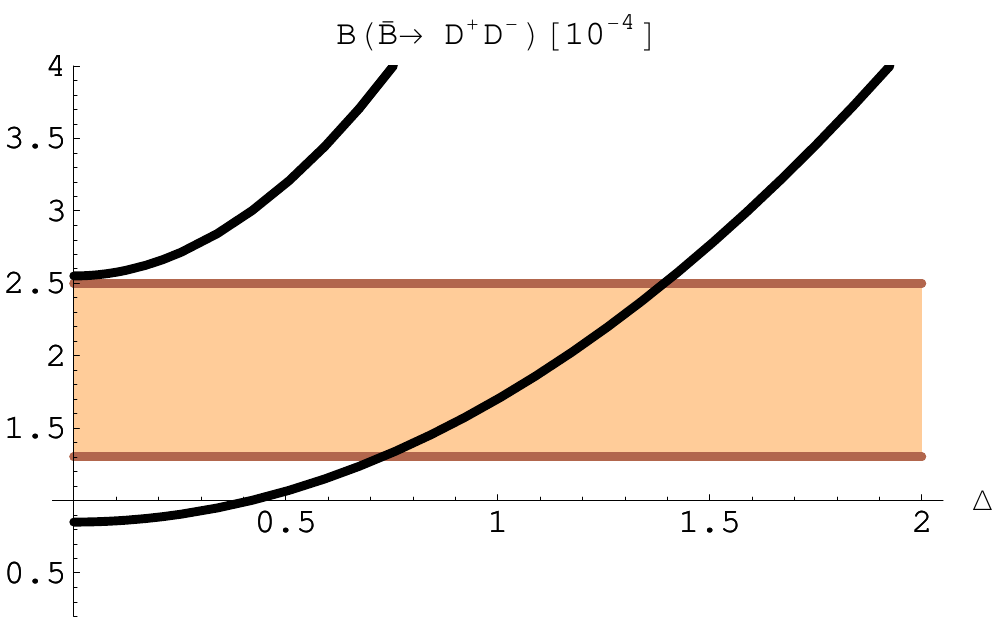}
 \includegraphics[width=2.2in]{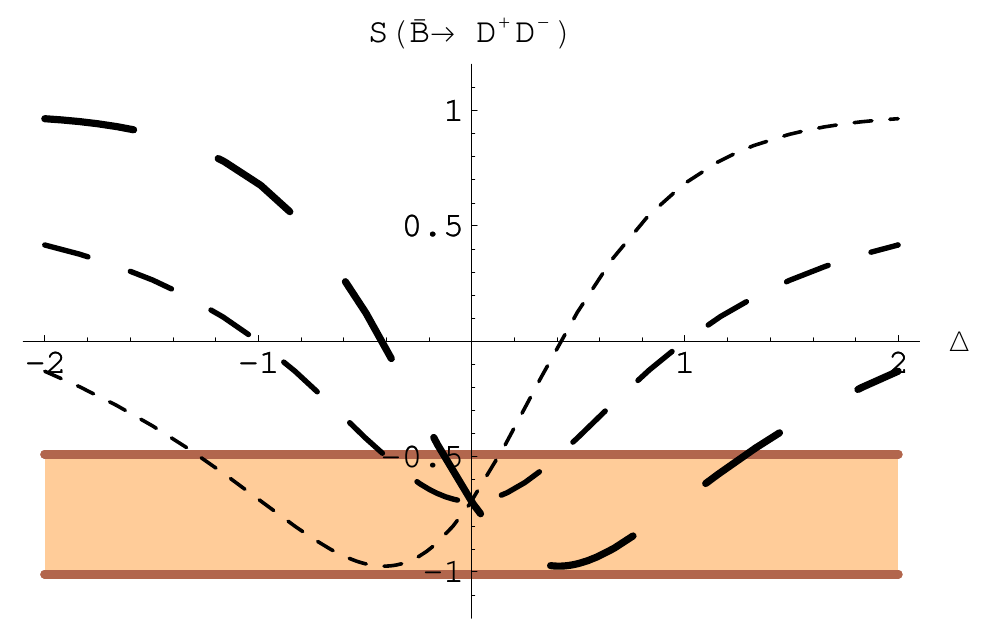}
  \includegraphics[width=2.2in]{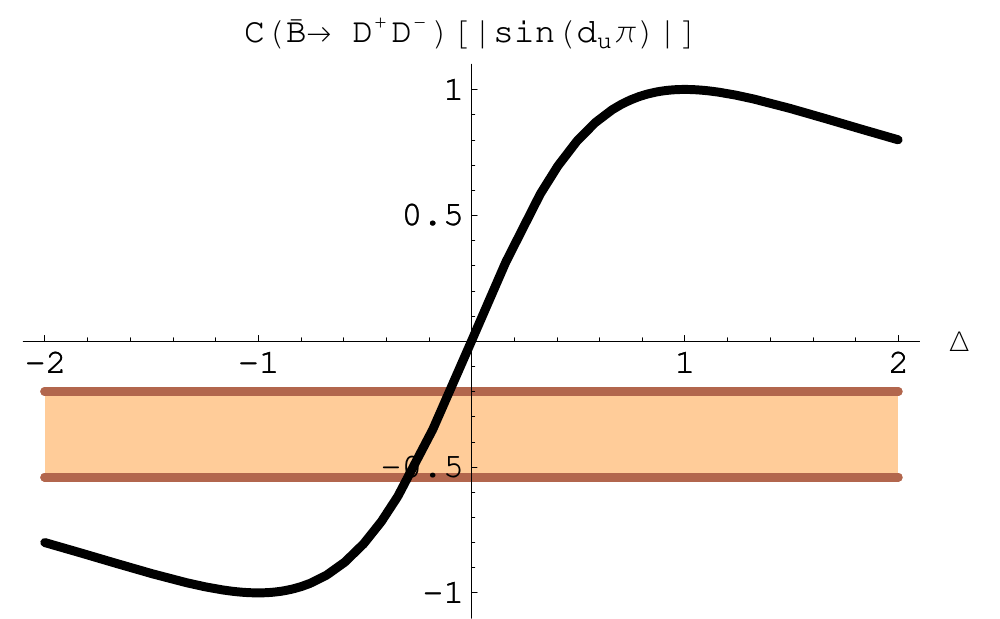}
 }
 \caption{\small  A weak phase difference $\phi = 90(270)^\circ$  is assumed here
 for $\Delta_{\rm DD}$ positive (negative). 
 (left) Branching fraction \eqref{eq:BDD_analyze} as a function of $\Delta_{\rm DD}$.
 The black bands correspond to  the SM estimate
\eqref{eq:BDD_theory} at  $\Delta_{\rm DD}=0$. The brown-red band corresponds to the HFAG bounds
in Eq.~\eqref{eq:BDD_exp}. 
(middle) Time dependent CP asymmetry $S_{D^+D^-}$ as a function of  $\Delta_{\rm DD}$ for
$\dU = 1.1,1.5,1.9$ where the dashes get shorter for larger values of $\dU$. The interpolation between
those values is fairly smooth.
(right)  The CP asymmetry as a function of  $\Delta_{\rm DD}$ in units of $|\sin(\dU \pi)|$.    }
 \label{fig:BDD}
 \end{figure}
In order to look for maximal CP violation we may again set the weak phase
difference to $90(270)^\circ$  in the formulae in Eq.~\eqref{eq:BDD_analyze}.
In Fig.~\ref{fig:BDD} (left) the branching fraction is plotted as
a function of $\Delta_{\rm DD}$ with uncertainty taken from the SM estimate
\eqref{eq:BDD_theory} at  $\Delta_{\rm DD}=0$.
The brown-red band corresponds to the HFAG bounds
in Eq.~\eqref{eq:BDD_exp}. The new feature as compared to the $B \to \tau \nu$ analysis
is the constraint from $S_{D^+D^-}$ which corresponds to the figure in the middle.
The CP asymmetry is plotted to the right of that figure.
Once more the branching ratio does not set limits on the amounts of CP violation,
in fact the uncertainties are very similar as in $B \to \tau \nu$. 
Demanding the uncertainty bands to be tangential at worst results in
$|\Delta_{\rm DD}| < 1.5$.
The constraints from $S_{D^+D^-}$  do depend
on the scaling dimension. The parameter $\dU = 1.1$ for example seems
slightly disfavoured as compared to the value $\dU =1.9$  

\subsubsection*{Weak phase $\phi \neq 90(270)^\circ$}

\begin{figure}[!p]
 \centerline{\includegraphics[width=2.4in]{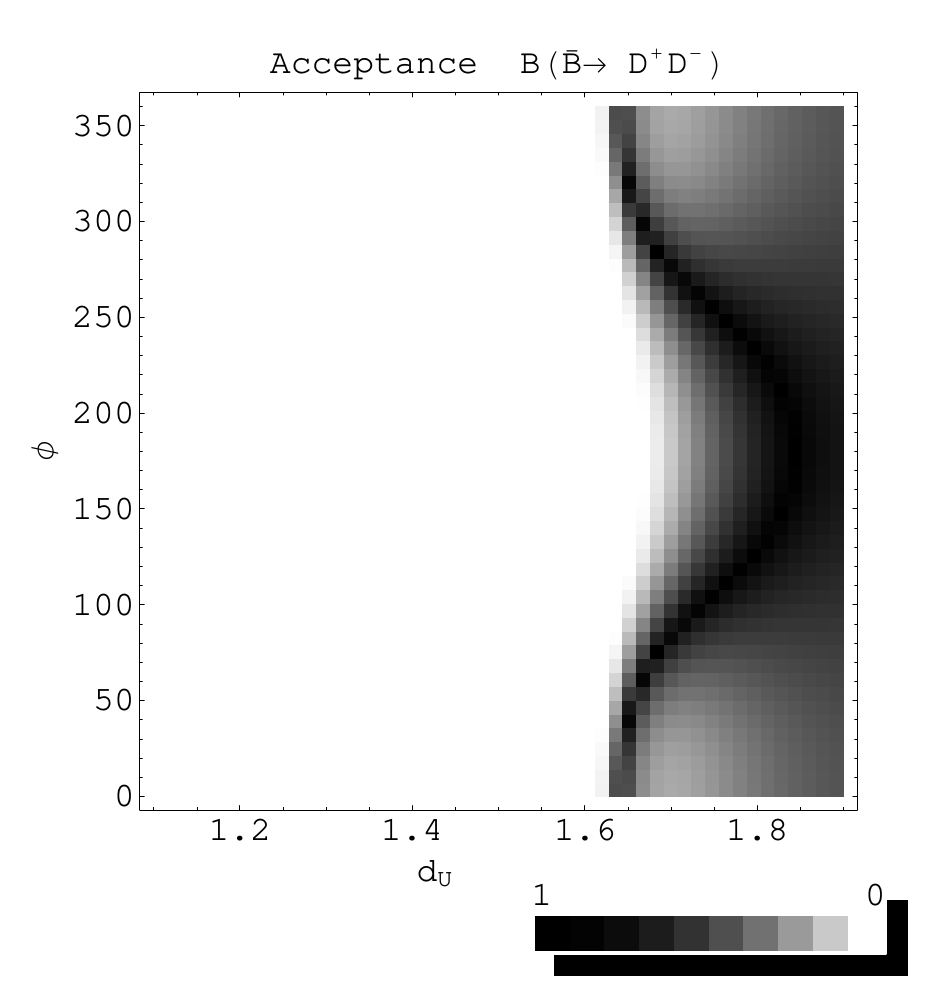},
 \includegraphics[width=2.4in]{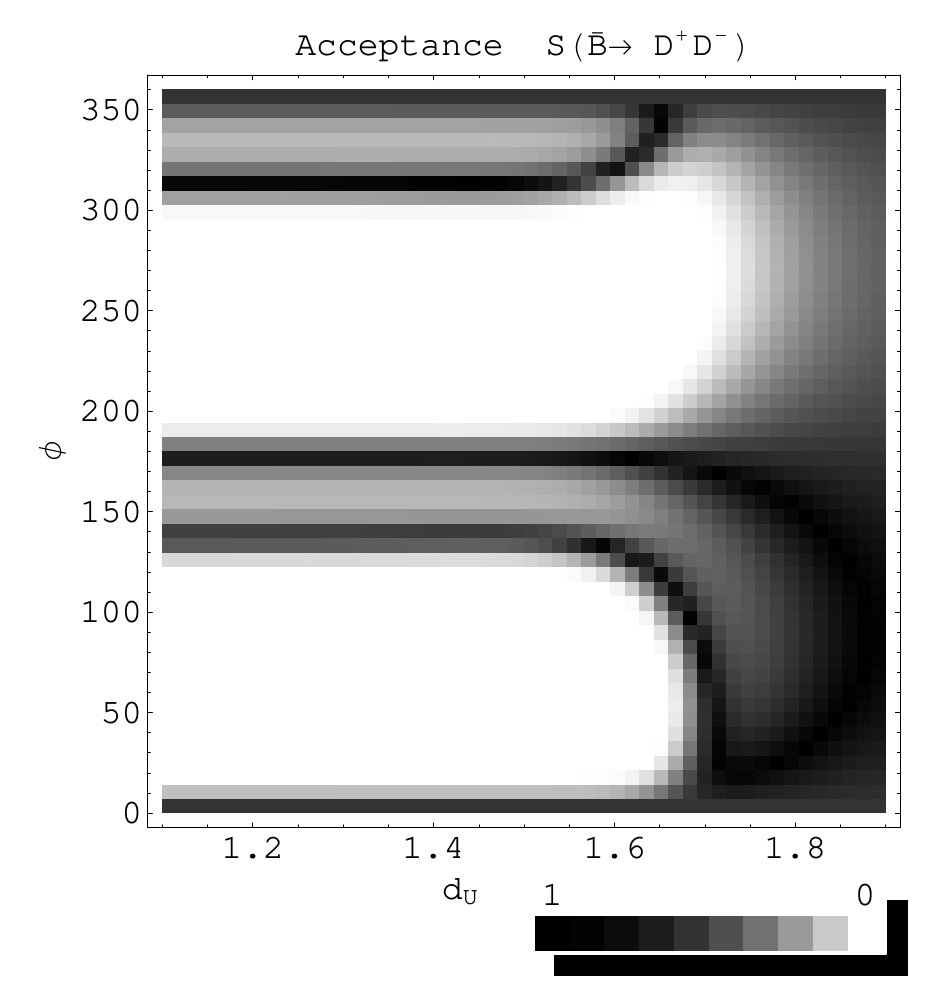},
 \includegraphics[width=2.4in]{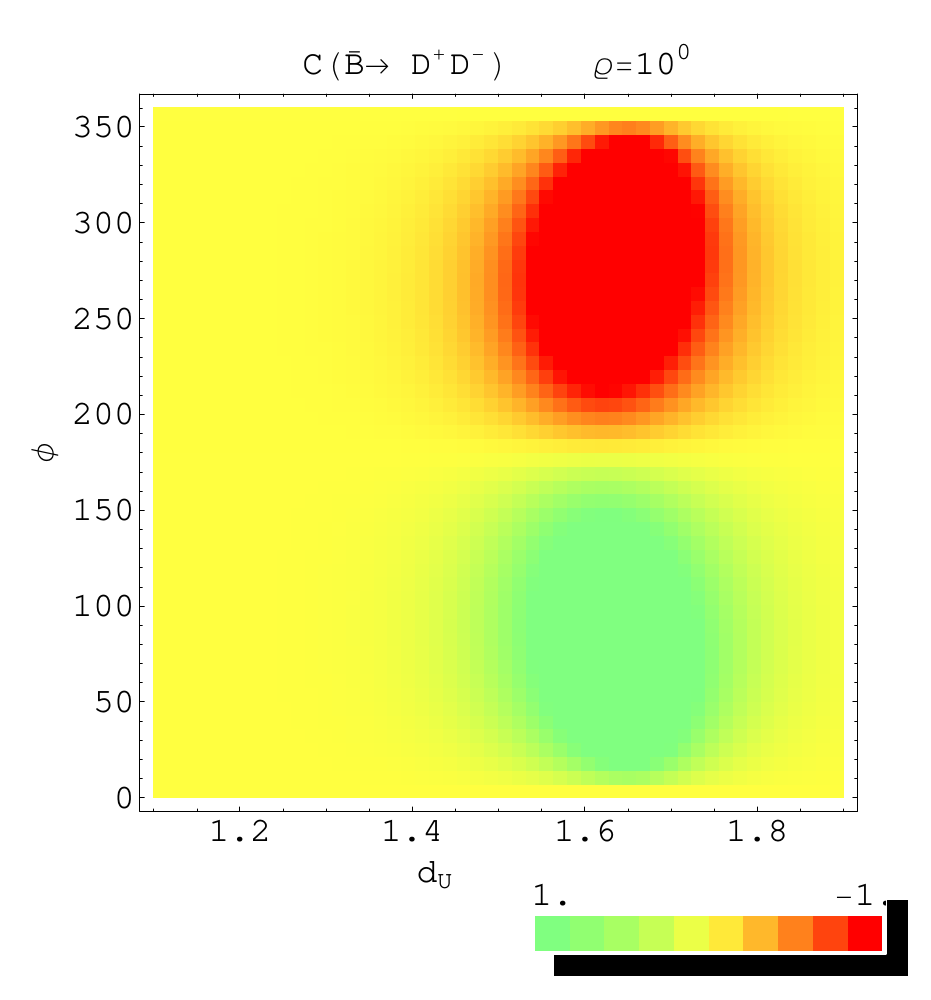}}
 \centerline{\includegraphics[width=2.4in]{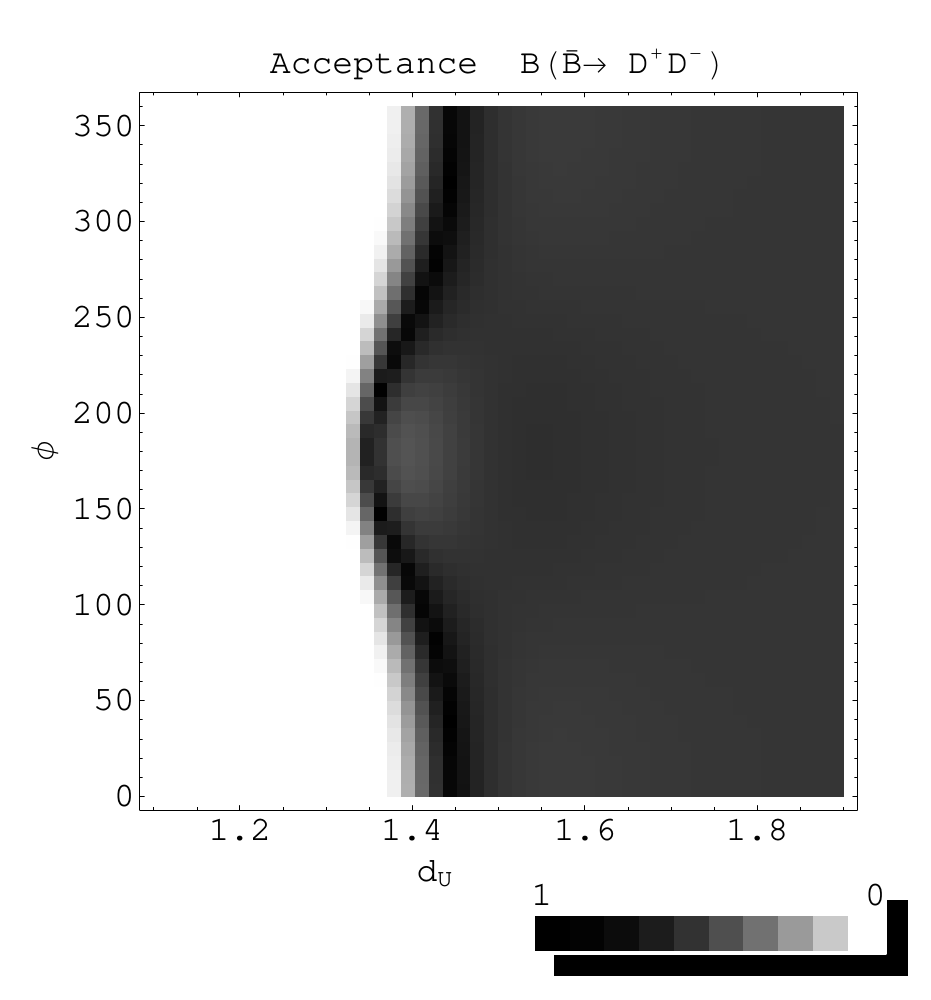},
 \includegraphics[width=2.4in]{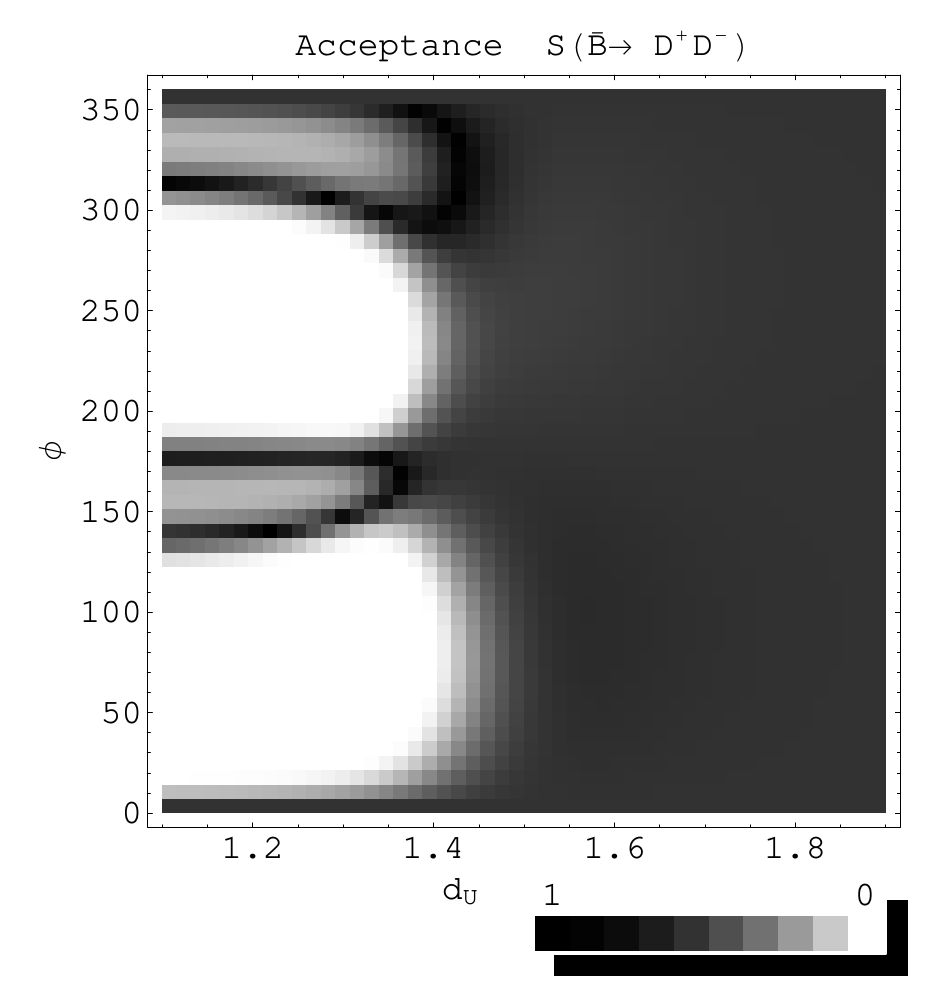},
 \includegraphics[width=2.4in]{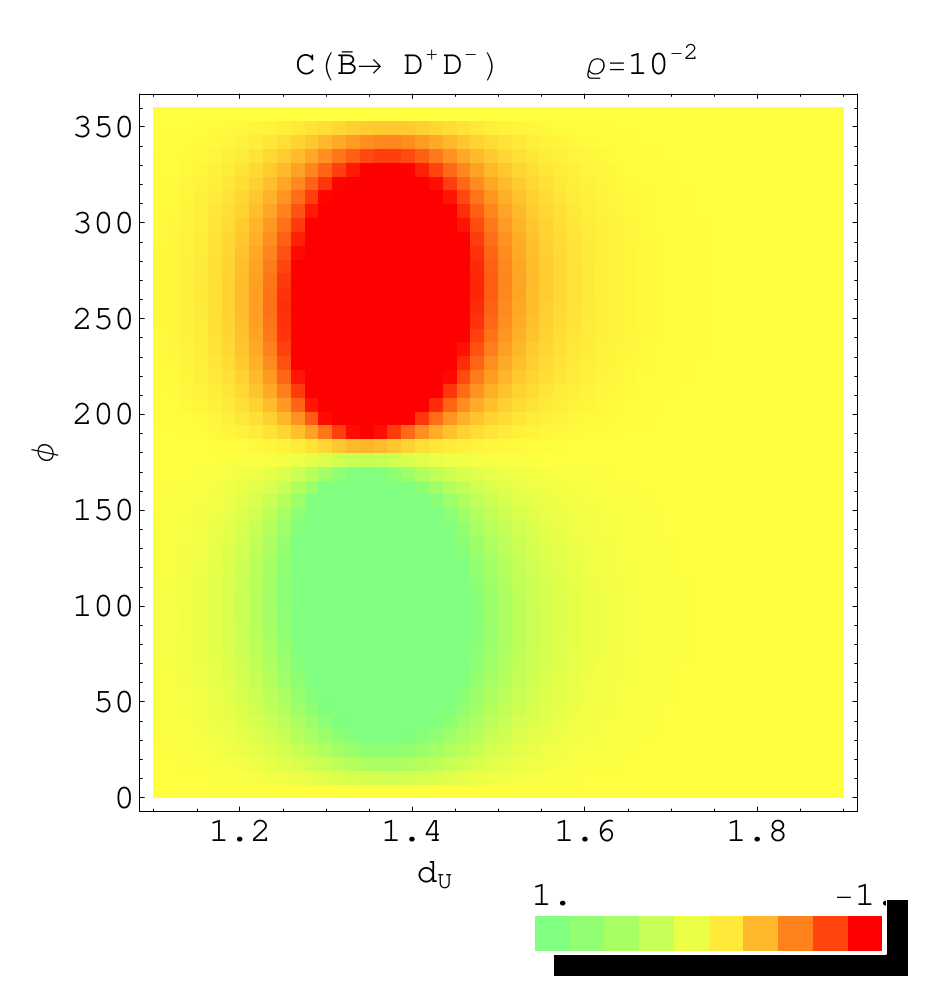}}
 \centerline{\includegraphics[width=2.4in]{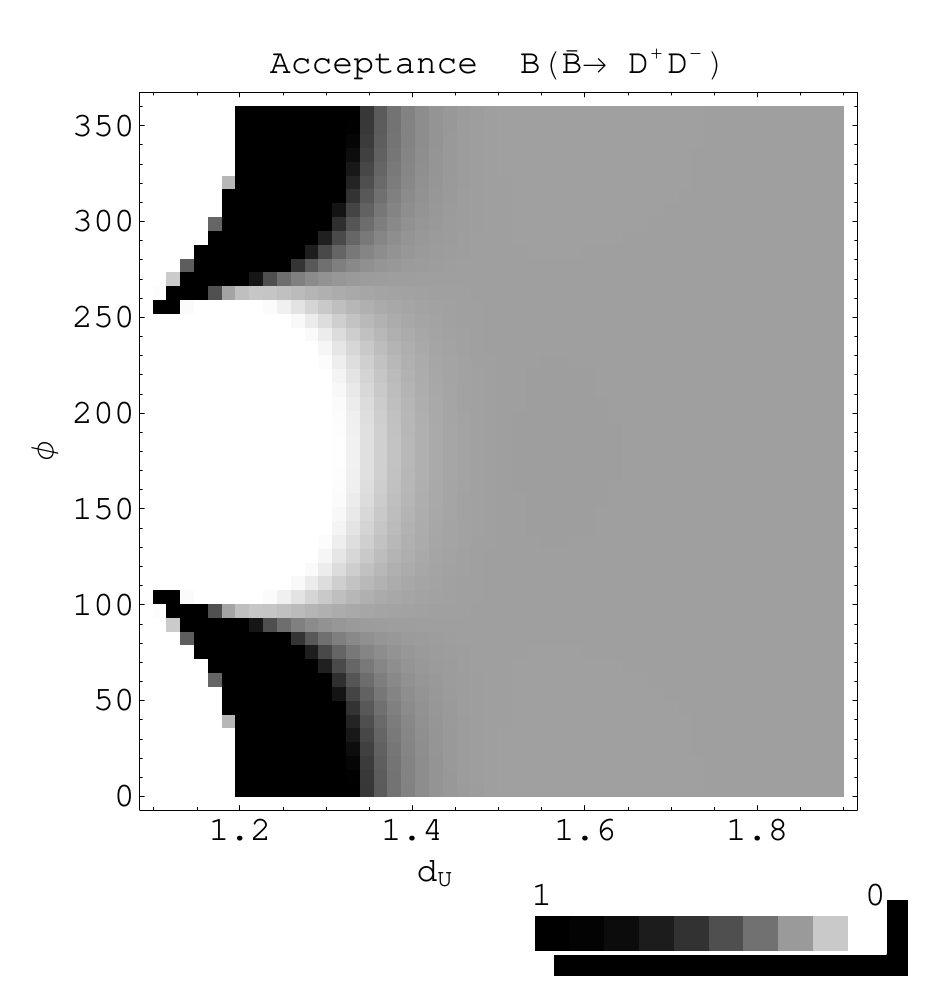},
 \includegraphics[width=2.4in]{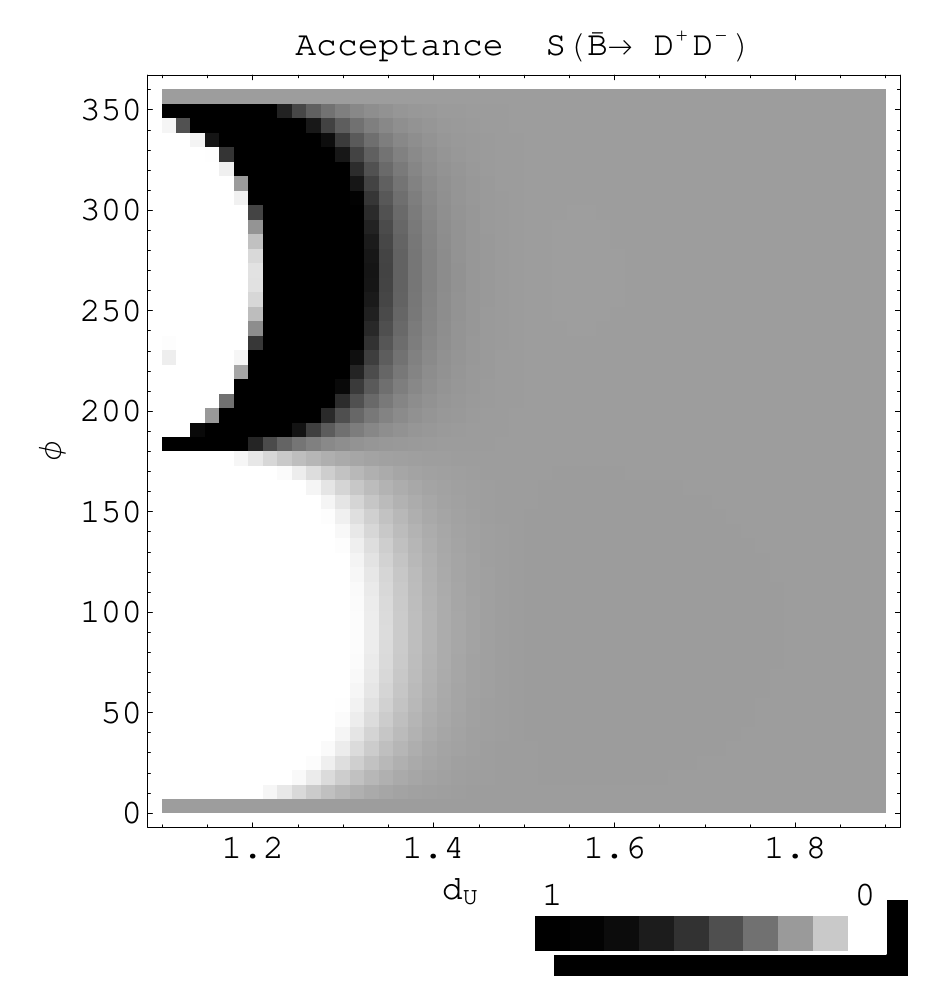},
 \includegraphics[width=2.4in]{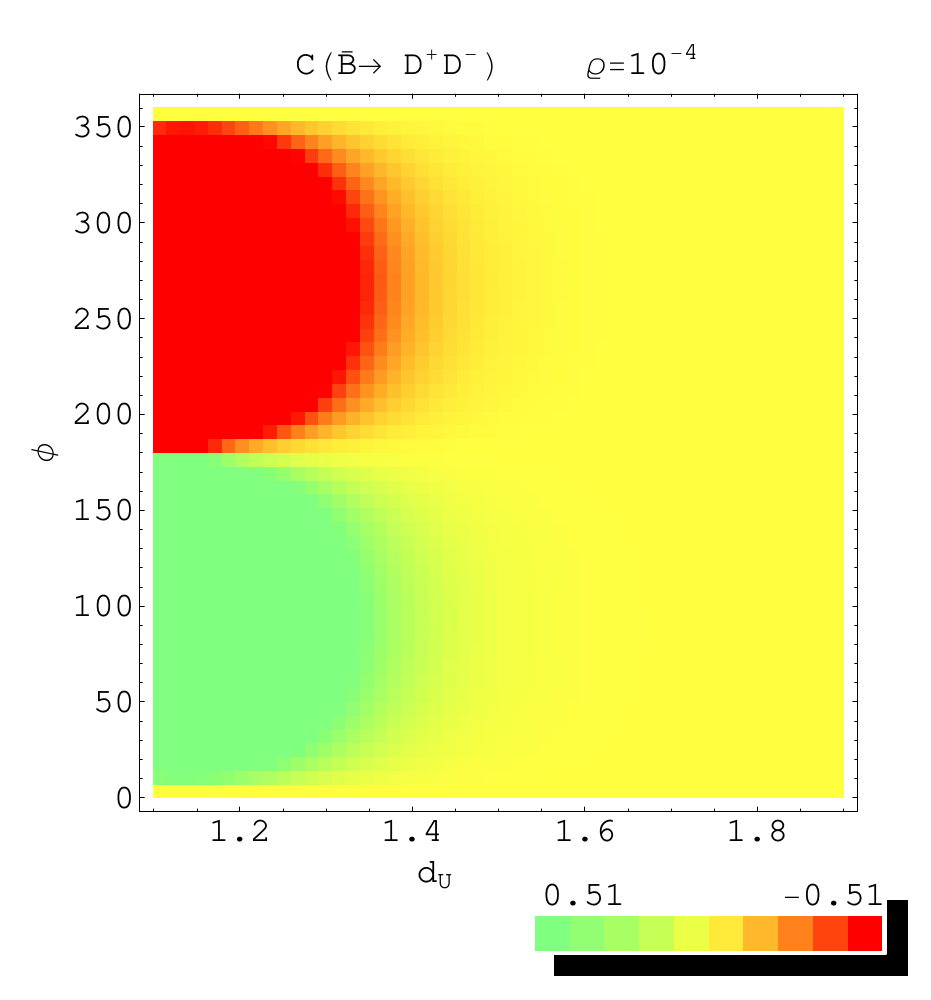}}
  \caption{\small The  observables with fractions of effective couplings
 $\rho = 10^{0,-2,-4}$, as defined in \eqref{eq:rhoBDD},  are plotted from the top of the figure to the bottom.
 Constraints on the $(\phi,\dU)$ parameter-space from (left) the branching fraction  and (middle) the CP asymmetry $S_{D^+D^-}$ (middle). 
The values in the  dark regions are allowed whereas white ones are disfavoured. c .f. text for more details.
 (right) The CP asymmetry $C_{D^+D^-}$  as a function of  $(\phi,\dU)$.
 The scale $\Lambda_\cU = 1\,{\rm TeV}$ is chosen here.}
 \label{fig:BDDCP}
 \end{figure}

We investigate the  two dimensional  parameter space $(\phi,\dU)$  for 
different ratios of effective couplings.
These quantities relate to  $\Delta_{\rm DD}$ \eqref{eq:deltaDD} as follows
\begin{eqnarray*}
\label{eq:delDDex}
\Delta_{\rm DD}  &=&  \rho_{\rm DD}
  \frac{A_\dU}{2 \sin(\dU \pi)} \frac{m_D^2}{m_c (m_b-m_c)}    \,
\frac{ (G_F/\sqrt{2})^{-1} }{m_D^2} \Big( \frac{m_D^2}{\Lambda_\cU^2}  \Big)^{\dU-1}    \\
&\simeq&  17\cdot10^{3} \Big( 3.5 \cdot10^{-6}  \frac{\Lambda_\cU}{1\,{\rm TeV}}  \Big)^{\dU-1} \frac{A_\dU}{\sin(\dU \pi)}  \rho_{\rm DD} \,,
\end{eqnarray*}
where
\begin{equation}
\label{eq:rhoBDD}
 \rho_{\rm DD} \equiv \frac{|\lambda_{S}^{\rm cb}\lambda_{P}^{\rm cd} |}{|V_{\rm cb} U_{\rm cd}|} \,.
\end{equation}
A plot relating $\Delta_{\rm DD}$ and $\dU$ can be found in appendix
\ref{app:adu}, Fig.~\ref{fig:deltas}.
In Fig.~\ref{fig:BDDCP} (right) CP asymmetry $C_{D^+D^-}$ 
is plotted as a function $(\phi,\dU)$ for $\rho_{\rm DD} = (10^{0,-2,-4})$.
The pattern is very similar in its form to $B \to \tau \nu$.
A large asymmetry is obtained for $|\Delta_{\tau\nu}| \sim 1$, which 
cannot be attained for smaller values $\rho_{\rm DD}$.
The constraint on the branching fraction, Fig.~\ref{fig:btauBC} (left), and the CP asymmetry $S_{D^+D^-}$
are evaluated with the same kind of acceptance function as for $B \to \tau \nu$ \eqref{eq:defi}. 
The corresponding values for the CP asymmetry are
$S^{\rm SM}_{D^+D^-} = - \sin(2 \beta) = 0.69$, $S^{\rm HFAG} = -0.75$ and $\Delta S = 0.52$ 
corresponds to two standard deviations. The values for the branching fraction are 
${\cal B}^{\rm SM}_{D^+D^-} = 1.7 \cdot 10^{-4}$, ${\cal B}^{\rm HFAG}_{D^+D^-} = 1.9 \cdot 10^{-4}$  and
$\Delta {\cal B} = 1.6$ corresponds to linear addition of the theoretical and experimental uncertainty.

A qualitative result that can be inferred from Fig.~\ref{fig:btauBC} is that the parameter
space of a large positive CP asymmetry $C_{D^+D^-}$ is disfavoured by the bounds from the
$S_{D^+D^-}$. This is easily seen from the formulae \eqref{eq:BDD_analyze}, \eqref{eq:deltaDD}
and the plots in appendix \eqref{app:adu}. A negative $C_{D^+D^-}$ demands a weak phase $\phi < 180^\circ$
and then the linear and quadratic terms in $S_{D^+D^-}$ add constructively and are in conflict with
the consistent result between the SM and experiment in this observable.
As for $B \to \tau \nu$ for small $\rho_{\rm DD}$ the linear terms dominate the 
quadratic ones and a regular pattern in $\cos(\phi)$ and $\sin(2\beta-\phi)$ emerges.

 \subsection{Discussion of $B_d \to D^+D^-$
 and remarks on U-spin \& colour related channels}
 \label{sec:disBDD}

A large  CP asymmetry $C_{D^+D^-}$ would be a rather puzzling fact,
as for instance discussed in Ref.~\cite{Ikaros}.
One is lead  to suspect that the gluonic penguin 
$B_d \to D \bar q q$ with $q = c$ might be enhanced by new physics.
This scenario would or should  lead to enhanced penguin amplitudes
for $q = (u,d,s)$ as well and  enter
$B_d \to (\pi\pi,K K)$ in disagreement with
the $B$-factory data.

We have seen that an unparticle scenario can lead, for
appropriate parameters, to enhanced CP violation.
One might wonder whether similar results
shouldn't  also show up in U-spin ($ s \leftrightarrow d$) and colour related channels. The plots in Fig.~\ref{fig:BDDCP} indicate
that the CP asymmertry $S$ in general does not necessarily receive 
large contributions. This can be inferred from 
Eq.~\eqref{eq:BDD_analyze} or by noting that the unparticles
just contribute to a large SM background from 
$\sin(2 \beta)$. We shall therefore focus on the 
CP asymmerty C. Let us note however that 
the situation for $B_s$ decays is 
different since the mixing phase $\phi_s \simeq 0$ 
($\phi_d \simeq 2\beta$) in the 
SM and the contributions of unparticles would be 
not be shielded by a large SM value.

The colour related or colour suppressed channel of 
$B_d \to D^+D^-$ is $B \to J/\Psi \pi_0$. The CP asymmetry 
has been measured $C_{J/\Psi \pi_0} = -0.11(20)$ \cite{PDG},
which is not conclusive in regard to its size. 
In the colour suppressed modes the non-factorizable 
contributions are enhanced due to different combinations 
of Wilson coefficients  (typically 
$\sim 2\!-\!3$ larger than the factorizable amplitude) and have
large strong phases.
On the practical side it is harder to estimate them 
reliably in the SM and even more in the unparticle scenario, where
the unparticle is dynamical as compared to the
contracted $W$-boson propagator in the SM.
The strong phases and the  different hierarchy 
between factorizable and non-factorizable contributions 
in the SM and the unparticle scenario\footnote{A parametric estimate gives that the non-factorizable
contributions in the unparticle scenario are suppressed 
by a factor $2m_D^2/(m_{J/\Psi}^2 + m_B^2) \sim 0.2$ as compared to the SM.} make it impossible to draw conclusions without explicit 
calculations.

The U-spin related transitions $b \to \bar c c s$ are CKM enhanced
and therefore statistics should make them more attractive. 
In principle there is no reason that generic new physics respects
the CKM hierarchy and U-spin. In the unparticle scenario 
there is no principle that dictates a CKM-like hierarchy
in the coefficients $\lambda_{q'q}$ in the effective Lagrangian 
\eqref{eq:leff}. Therefore they are not necessarily of 
major concern. Let us nevertheless discuss them.
The gold plated decay $B \to J/\Psi K_s$ is also colour-suppressed.
The measurement  of the CP asymmetry $S_{J/\Psi K_s}  = \sin(2\beta)$ has allowed determination of 
the angle $\beta$ in the SM,
whereas the CP asymmetry $C_{J/\Psi K_s} = 0$ is consistent with experiment.
This mode is highly consistent with the SM or more precisely with one dominant amplitude.
The branching fraction of the colour allowed  decay 
$B_d \to D^+ D_s^-$ has been measured 
but no CP asymmetry has been reported, presumably because it does not exhibit CP violation in mixing. 
If the Belle CP asymmetry in  $C_{D^+D^-}$ gets confirmed a look at the CP asymmetry
appears mandatory. 

In summary the most interesting parallel channel is 
probably $B_d \to J/\Psi\pi0$ and the improvement
of the measurement in $C_{J/\Psi\pi_0}$ should be
watched along with $C_{D^+D^-}$. 
In the scenario we described we would generically
expect a large CP asymmetry $C_{D^+D^-}$ to 
be accompanied  by a large asymmetry in $C_{J/\Psi \pi_0}$.
It is a serious point of criticism,
but on the other hand the experimental result is not conclusive
and in theory there might be cancellations between the strong
phase $e^{i\dU\pi}$ and the phase from the non-factorizable 
interactions.
The time dependent
CP asymmetries $S$ are shielded by large SM 
backgrounds for $B_d$-meson, whereas in $B_s$
system the SM expectation is $S \sim 0$ in many 
cases (e.g. $B_s \to J/\Psi \phi$) and the unparticle 
scenario might reveal itself. 

We have seen that CP violation in $B_d \to D^+D^-$
and $B \to \tau \nu$ can be maximal in the unparticle
scenario. After this phenomenological section we shall 
elaborate on whether a CP asymmetry in leptonic decays 
is possible.  Thereafter we shall  
turn to the question of whether
the scale invariance at the ${\rm TeV}$-scale or near
scale-invariance could still be effective at heavy flavour 
scales $\sim 5\,{\rm GeV}$.

\section{Constraints from CPT on (new) CP-violation}
\label{sec:cpt}

The invariance under  CPT symmetry imposes constraints on 
the amount of CP-violation;  it enforces the 
equality of the partial sum of rates of particles and antiparticles\footnote{I am grateful to Ikaros Bigi for drawing my 
attention to this fact.},
to be made more precise below.
Neither the SM nor any well-known new physics model 
predict CP-violation in leptonic decays such as
$B \to \tau \nu$ studied in this paper. The aim of this section
is to verify explicitly whether the CP-violation is consistent with the 
constraints from CPT. 

Let us note that we expect that CPT-invariance
holds for a theory with a local hermitian Lagrangian 
such as in Eq.~\eqref{eq:leff}. The explicit verification of CPT invariance
demands that $\Theta {\cal L}(x) \Theta^{-1} = {\cal L}^\dagger(-x) = {\cal L}(-x)$, where $\Theta = CPT$ denotes the combined 
CPT-transformation.
The Lagrangian \eqref{eq:leff} fulfills this requirement provided that
$ \Theta O_\cU (x) \Theta^{-1} = O_\cU ^\dagger(-x)$, which we 
cannot verify explicitly since we do not have equations of motions  or a Lagrangian for the unparticle field at hand from where we would infer the transformation under $C$, $P$ and $T$. 
There also exists  a general proof of the CPT-theorem in the
framework of axiomatic field theory \cite{Jost} based on 
general principles and axioms such as
Lorentz invariance, uniqueness of the vacuum and
causality of field commutators. 
Concerning the latter we would like to mention that we 
have seen in a previous section that the unparticle
field obeys  causality, c.f. Eq.~\eqref{eq:comm1}.
Summarising, although we are not able to prove CPT-invariance 
we at the same time do not find any indications why
it should be violated.

It is well known that
CPT symmetry implies equality of the decay rates of particles
and antiparticles. In practice there
is even a  stronger consequence 
, e.g. \cite{wolf}, \cite{bigisanda} or \cite{GerardHou} where it was applied to charmless $B$-decays.
The final state particles can be divided into subclasses
of particles which rescatter into each other.
It is a fact that  the sum of the
partial rates of these subclasses for a particle and
its antiparticle must be the same. This can be inferred
from the following relationship \cite{wolf}
between the weak decay amplitudes of a $B$-meson and
its antiparticle $\bar B$ to a final state $f_x$
\begin{equation}
\label{eq:CPT}
\matel{\bar f_x}{H_{\rm decay}}{\bar B}^* = 
\sum_{i} \matel{f_x}{S^\dagger}{f_i}\matel{f_i}{H_{\rm decay}}{B} \,,
\end{equation}
where $H_{\rm decay}$ corresponds to the weak transition operator
and $S$ is the scattering matrix.
This relation is derived from the completeness relation 
${\bf 1} =  \sum_{i}\state{f_i}\astate{f_i}$  and the fact that the CPT-operator is antiunitary. An equivalent but alternative relation on the level of decay rates can be found in Ref.~\cite{bigisanda}.
From Eq.~\eqref{eq:CPT} it is then inferred that all states $f_j$ which
rescatter into $f_x$ form a subclass whose partial rates of particles
and antiparticles sum to zero
\begin{eqnarray}
\label{eq:genCPT}
\sum_{i \in I} \Delta \Gamma(B \to f_i) = 0 \,, \qquad 
\matel{f_i}{S^\dagger}{f_j} \neq 0 \quad i,j \in I \,,
\end{eqnarray}
where
\begin{eqnarray}
\Delta \Gamma(B \to f)  \equiv \Gamma(B \to f)-
\Gamma(\bar B \to \bar f) \quad .
\end{eqnarray}

\begin{figure}[h]
 \centerline{\includegraphics[width=5.2in]{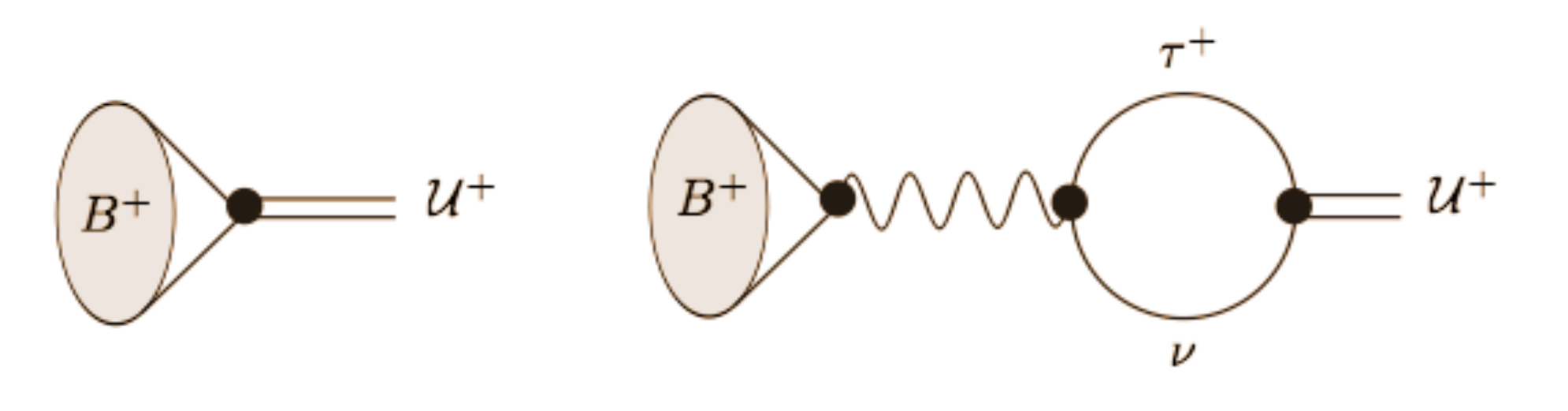}}
 \caption{\small Decay $B^+ \to \cU^+$, the double lines denote 
 an unparticle (left)  leading order (right)
 with virtual $\tau\nu$-loop correction .}
 \label{fig:unparticle}
 \end{figure}

The exact relation between the CP asymmetry and the difference
of decay rates can be infered from Eq.~\eqref{eq:cpdef}.
Whereas the new CP asymmetry generated by 
${\cal A}_{\rm CP}(D^+D^-) \sim \Delta \Gamma( B_d \to D^+ D^-)$ may be compensated
by $ \Delta \Gamma( B_d \to \bar D_0  D_0)$ for instance,
it is at first sight not clear which mode would compensate for the
new CP asymmetry in ${\cal A}_{\rm CP}(\tau \nu) \sim \Delta \Gamma( B^+ \to \tau^+ \nu)$.
Among the SM final states there does not seem to be an 
appropriate candidate. We are led to look in the unparticle 
sector for a suitable candidate. A firm hint can be
gained by counting the coupling constants.
Denoting the weak coupling by $v$ and 
the unparticle coupling by $\lambda$ \eqref{eq:leff}, the 
CP asymmetry, which arises  due to an interference of the
two amplitudes depicted in Fig.~\ref{fig:btau_feyn}, is of the order $O(\lambda^2 v^2)$.
The processes $B^+ \to \cal U^+$  with an interference of 
the two amplitudes depicted
in  Fig.~\ref{fig:unparticle} has the same counting in the coupling constants. One amplitude corresponds to a tree decay and the 
other one incorporates a virtual correction due to a fermion loop
of the $\tau$ and the $\nu$.
The process $B^+ \to {\cal U}^+$ is kinematically
allowed since the unparticle has a continuous mass spectrum.
It does not proceed at resonance, but rather behaves like a
multiparticle final state and is a realisation of Georgi's observation that
the unparticle field in a final state 
behaves like a non-integral number $\dU$ of
massless particles.

We shall now explicitly verify the CPT constraint
\begin{equation}
\label{eq:CPTconstraint}
\Delta \Gamma(B^+ \to \tau^+ \nu) + 
\Delta \Gamma(B^+ \to {\cal U^+})_{\rm \tau \nu-loop} = 0 \,.
\end{equation}
For the sake of simplicity we shall assume  as  
previously that there is no flavour dependent perturbation in the
neutrino sector and that $\lambda_P^{\tau \nu} = - \lambda_S^{\tau \nu}$ in \eqref{eq:leff}. The formula for the first difference can be read off from
Eq.~\eqref{eq:btau_analyze}
\begin{eqnarray}
\label{eq:firstCPT}
\Delta \Gamma(B^+ \to \tau^+ \nu) &=& 
- 4 {\cal B}^{\rm SM}_{\tau \nu} \sin(\phi) \sin(\dU \pi) \Delta_{\tau \nu}
\\
& = & - \sin(\phi) \frac{G_F}{2\sqrt{2}\pi} \frac{m_B}{m_b} m_\tau 
f_B^2 \Big(1\!-\!\frac{m_\tau^2}{m_B^2}\Big)^2 |\lambda^S_{\tau \nu} \lambda^P_{ub} V_{\rm ub} U_{\tau \nu}| A_\dU 
\Big(\frac{m_B^2}{\Lambda_\cU^2}\Big)^{\dU-1} \nonumber 
\end{eqnarray}
Note that the cancellation of the phase factor $\sin(\dU \pi)$ by the
same factor in the denominator, as previously mentioned, is crucial
for the cancellation here since the graphs in Fig.~\ref{fig:unparticle} do not involve this factor!
The amplitude of the graph in  Fig.~\ref{fig:unparticle} to the left is
\begin{equation}
{\cal{A}}(B^+ \to {\cal U}^+)_{\rm Fig.\ref{fig:unparticle} (left)} 
=   \lambda_P^{\rm ub\,*}\, {\cal A}_1
=  \frac{\lambda_P^{\rm ub\,*}}{\Lambda^{\dU-1}} \frac{m_B^2}{m_b} f_B \matel{P}{O_\cU^\dagger}{0}
\end{equation}
and the amplitude of the graph to the right of Fig.~\ref{fig:unparticle}  
is 
\begin{eqnarray}
{\cal{A}}(B^+ \to {\cal U}^+)_{\rm Fig.\ref{fig:unparticle} (right)} 
&=&\lambda_S^{\rm \tau \nu \,*}  V_{\rm ub}^* U_{\rm \tau\nu} \, {\cal A}_2 \nonumber \\
&=&  \frac{\lambda_S^{\rm \tau \nu \,*}}{\Lambda^{\dU-1}} 
\frac{G_F}{\sqrt{2}} V_{\rm ub}^* U_{\rm \tau\nu} m_\tau  
f_B  \Pi_{S\!-\!P}(m_B^2) \matel{P}{O_\cU^\dagger}{0} \,
\end{eqnarray}
where we have factored the weak parameters in 
${\cal A}_{(1,2)}$.  The fermion-loop $\Pi_ {S\!-\!P} $ is given by the correlation function
\begin{equation}
\Pi_ {S\!-\!P}(p_B^2=m_B^2) = i \int d^4 x \, e^{-i p_B \cdot x} \,
\matel{0}{T 
[\bar \nu (1\!-\!\gamma_5)\tau](x)\, [\bar \tau (1\!-\!\gamma_5)\nu](0) }{0}
\,.
\end{equation}
The decay rate is calculated from
\begin{equation}
\Gamma = \frac{|{\cal A}|^2}{2 m_B} \int d\Phi \, \qquad  {\rm with} \quad \int d\Phi  = A_\dU (m_B^2)^{\dU-2}
\end{equation}
being the phase space volume.
The difference of decay rates is given by
\begin{equation}
\Delta \Gamma(B^+ \to \cU^+)_{\rm \tau \nu-loop} = 
4 \sin(\phi) {\rm Im}[{\cal A}_1^* {\cal A}_2]  A_\dU (m_B^2)^{\dU-2}
\end{equation}
Since ${\cal A}_1$ is real only the imaginary part of ${\cal A}_2$
  will enter.
The only strong phase is due to the $\tau$ and the $\nu$ going on-shell
in the loop in Fig.~\ref{fig:unparticle} (right).
Therefore we only need to know the imaginary part of the fermion loop
which   is given by
\begin{equation}
{\rm Im}[\Pi_ {S\!-\!P}(m_B^2+i0)] =   \frac{1}{4\pi } m_B^2 
\Big(1\!-\!\frac{m_\tau^2}{m_B^2}\Big)^2 \quad .
\end{equation}
Assembling the formulae we get
\begin{eqnarray}
\Delta \Gamma(B^+ \to \cU^+)_{\tau \nu} 
 =   \sin(\phi) \frac{G_F}{2\sqrt{2}\pi} \frac{m_B}{m_b} m_\tau 
f_B^2 \Big(1\!-\!\frac{m_\tau^2}{m_B^2}\Big)^2 |\lambda_S^{\tau \nu} \lambda_P^{ub} V_{\rm ub} U_{\tau \nu}| A_\dU 
\Big(\frac{m_B^2}{\Lambda_\cU^2}\Big)^{\dU-1} \,, 
\end{eqnarray}
which fulfills the CPT constraint Eq.~\eqref{eq:CPTconstraint}
together with \eqref{eq:firstCPT}.

We have explicitly verfied the CPT constraint \eqref{eq:genCPT}
for the decay $B \to \tau \nu$ with unparticle-SM interactions
given by the Lagrangian  \eqref{eq:leff}. We do not dare to speculate in any detail 
on how a decay $B^+ \to {\cal U}^+$ might be observed in a laboratory experiment. 
It can be said though that the unparticle has  directed momentum, mass and  charge which it directly inherits from the $B$-meson.
Moreover in the case where there is a CP asymmetry in 
$B \to \tau \nu$ 
due to unparticles, it is precisely  the CPT constraint
\eqref{eq:CPTconstraint} which tells us that there is 
an excess of charged unparticle degrees of freedom produced.
Whether a part of this charge could 
annihilate into neutral particles or decay into charged particles 
remains unclear since 
the nature of this degree of freedom remains unknown at this
stage. 
These questions could be addressed once a concrete model
realising the unparticle scenario is known.

\section{Breaking of scale invariance - dimensional analysis}
\label{sec:dim}

The SM at the electroweak scale is not scale invariant.
The logarithmic running  and in particular the vacuum expectation value 
of the Higgs, which give masses to the fundamental particles, are responsible
for the breaking of scale invariance. It is therefore a legitimate
question at what scale the symmetry breaking will be transmitted 
to the unparticle sector by the the effective Lagrangian
\eqref{eq:leff}.
This will depend on the strength of the coupling and 
the relevance of the operators in the latter.

The authors of  reference \cite{FRS} 
have  addressed this question, which we shall adapt  accordingly for the weak sector.
Assuming the unparticle field couples to an operator 
acquiring  a definite mass scale, the latter will break scale
invariance at some energy.
Let us assume for instance that 
the Higgs  couples to the unparticle operator,
following Ref.~\cite{FRS}, as follows
\begin{equation}
\label{eq:higgseff}
{\cal L}^{\rm eff} = \frac{ \lambda_H}{\Lambda_\cU^{d_{\cU_0}-2}  } |H|^2 O_{\cU_0}
\end{equation}
with  $\lambda_H = c^H_\cU (\Lambda_\cU / M_\cU)^{ d_{{UV}_0}-2}$ in our notation. We have used a new symbol $O_{\cU_0}$ for the unparticle
operator. This operator is not the same as the one used in 
Eq.~\eqref{eq:leff} since it has to be electrically neutral. The important question
for the analysis in this paragraph is what the value  the anomalous dimension 
$\bar d_{\cU}$ assumes.  
In the case where we think of the unparticle as being charged under 
$SU(2)_{ L}$, $O_{\cU_0}$ would appear
as  $ \delta {\cal L}^{\rm eff} \sim \bar q (\gamma_5) q O_{\cU_0}$ in addition
to the effective Lagrangian \eqref{eq:leff}
and $d_{\cU_0} = \dU$ seems unavoidable.
In the case where $O_\cU$ is the only unparticle field then 
$O_{\cU_0} = O_\cU O_\cU^\dagger$ would be a composite field
with anomalous dimension  in the range $0 \leq d_{\cU_0} \leq 2 d_\cU$,
where the Thirring model at coupling $\lambda = 2\pi$ \cite{Wilson} would be an example saturating the lower bound and supersymmetric QCD at the
conformal IR fixpoint
\cite{seiberg_intriligator}
an example saturating the upper bound. In the following we shall quote
values for the bounds and the mean value explicitly.
The Higgs VEV $\aver{ |H|^2 } = v^2$  is expected to break scale invariance
at a  scale $\tilde \Lambda$
\begin{equation}
 \frac{\lambda_H}{\Lambda_\cU^{d_{\cU_0}-2}} v^2 
 \tilde \Lambda^{d_{\cU_0}}  
= \tilde \Lambda^4  \qquad \Rightarrow \quad  \tilde \Lambda = \Lambda_\cU (\lambda_H \frac{v^2}{\Lambda_\cU^2})^{\frac{1}{4-d_{\cU_0}}}   \, .
\end{equation}
What would this scale be in the cases we have investigated?
Besides  $\Lambda_\cU$ there are two unknowns in the equation
above, first $d_\cU$ which appears explicitly in our results
and $\lambda_H = c^H_\cU (\Lambda_\cU / M_\cU)^{ d_{{UV}_0}-2}$.
In the latter the matching coefficient will remain unknown but
we can extract the ratio $(\Lambda_\cU/M_\cU)$
from  $\rho_{DD}(\rho_{\tau \nu})$ in 
terms of the UV dimensions.
Taking $B \to DD$ as an example the breaking scale is
\begin{equation}
\tilde \Lambda = \Lambda_\cU \Big(  c_\cU^H \frac{v^2}{\Lambda_\cU^2}
\big( \frac{\rho_{\rm DD}}{ R_{\rm DD} }\big)^{\frac{d_{{UV}_0}-2}{2(d_{UV}-1)}} \Big)^{\frac{1}{4-d_{\cU_0}}} \quad ,
\end{equation}
where $R_{\rm DD} = |c_S^{\rm cb} c_S^{\rm cd}|/|V_{\rm cb}V_{\rm cd}| = 1$,
deviating it from $1$ corresponds to a readjustment of $\rho_{\rm DD}$ in
terms of the matching coefficients. 
Assuming  for example $\Lambda_U = 1\,{\rm TeV}$, $\dU = 1.2$, 
$d_{\cU_0} = (0,1.2,2.4)$,  $\rho_{\rm DD} = 10^{-3.5}$, the ratio
of amplitudes and the breaking scale for fixed values of UV dimensions
become
\begin{alignat}{2}
& \Delta_{\rm DD} \simeq -0.40 R_{\rm DD} & &   \, \nonumber \\
& (d_{UV},d_{{UV}_0}) =  (3,6) \qquad & & 
 \tilde \Lambda  \simeq (67,20,1)~{\rm GeV}  (R_{\rm DD} c_\cU^H)^{1/(4.0,2.8,1.6)} 
\nonumber \\
& (d_{UV},d_{{UV}_0}) =  (3,3) \qquad & & 
 \tilde \Lambda  \simeq (300,180,50)~{\rm GeV}  (R_{\rm DD}^{-1/4} c_\cU^H)^{1/(4.0,2.8,1.6)} \quad . 
\end{alignat}
The situation is not conclusive, which is not surprising bearing in mind
that in the absence of a model there are simply to many unknowns.
In the case where both UV dimensions are the same, which should
be the case when $O_\cU$ and $O_{\cU_0}$ result from the same 
structure, a small matching coefficient $c_\cU^H$ is needed for
a sizable effect at the heavy flavour scales. If the UV dimensions
differ by a factor of two, which is the case when 
$O_{\cU_0} = O_\cU O_\cU^\dagger$, effects are possible for
moderate matching coefficient $c_\cU^H$ .

The effect $\Delta_{\rm DD} = -0.40$ appears  larger than the analysis or conclusions
in Ref.~\cite{FRS} suggest. There are two reasons. First and simply,
the CP violating phenomenon investigated in this paper is linear in the ratio of amplitudes,
whereas \cite{FRS} describes a case where the effect is proportional to the square
of the amplitude.
Secondly it was assumed that the SM Lagrangian has dimension four.
The crucial point is that the weak Lagrangian has 
dimension six, $d_{{\cal L}_{\rm weak}} = 6$ being suppressed by two powers of the
weak scale, whereas the unparticle Lagrangian 
has dimension $d_{{\cal L}_{\rm unp}} = \dU + d_{\rm SM}$.
In terms of the effective Lagrangian \eqref{eq:leff} and the Eq.~\eqref{eq:d12},
$4 < d_{{\cal L}_{\rm unp}} < 5$  the unparticle operator is more relevant
than the weak operator. 
This gives rise to an enhancement factor in the amplitudes
\begin{equation}
\frac{(G_F/\sqrt{2})^{-1}}{\mu_{\rm HF}^2} = \frac{8 m_W^2}{g^2 \mu_{\rm HF}^2} \, ,
\end{equation}
which is explicit in the results of Eqs.~\eqref{eq:deltatau}, \eqref{eq:deltaDD}.
In more physical terms one could state that  the weak boson propagates at the high weak scale whereas
the unparticle propagates at the low heavy flavour scale.

Adapting the analysis of Ref.~\cite{FRS} we imagine an experiment
at a scale $\mu_{\rm HF}$, the unparticle Lagrangian \eqref{eq:leff}
scales as ${\cal L} _{\rm eff} = \lambda_S/\Lambda_\cU^{\dU + (d_{\rm SM} -4)} \mu_{\rm HF}^{d_{\rm SM} + \dU}$,
the weak Lagrangian  as
${\cal L}_{\rm weak} \simeq G_F \mu_{\rm HF}^6$ and the ratio is
\begin{equation}
\Delta
\simeq   \lambda^S_\cU  \Big( \frac{\Lambda_\cU}{M_\cU} \Big)^{d_{UV}-\dU} 
\Big( \frac{\mu_{\rm HF} }{M_\cU} \Big)^{\dU + d_{\rm SM}-6} 
\Big( \frac{G_F^{-1} }{\mu_{\rm HF}^2} \Big) \,.
\end{equation}
Imposing that the energy scale
of the experiment is higher than the breaking scale,
i.e. $\mu_{\rm HF} > \tilde \Lambda$ ,
the following bound is obtained\footnote{Setting $c_\cU \to 1$, the fourth and the fifth term to one and taking the square root of the equation,
the bound in Ref.~\cite{FRS} is recovered with $\Delta^2 = \epsilon$.}
\begin{equation}
 \Delta <   \frac{c^S_\cU}{c^H_\cU} \, \Big( \frac{\mu_{\rm HF} }{M_\cU} \Big)^{d_{\rm SM}-2} 
  \Big( \frac{\mu_{\rm HF}^2 }{v^2 } \Big)  \Big( \frac{G_F^{-1}} {\mu_{\rm HF}^2 } \Big) \,   \Big( \frac{\mu_{\rm HF}}{\Lambda_\cU}  \Big)^{\dU - d_{\cU_0}} \quad .
\end{equation}
This equation is easily interpreted. The first factor measures the ratio of the two couplings.
The second is a measure between the relevance or dimension of the SM operator that is coupled to the unparticle 
and the dimension of the Higgs operator. In the third
term the scale of the experiment has to compete with the Higgs VEV. 
The fourth term is peculiar to the weak interactions, as described above, and is due 
to the fact that the weak process takes place at the weak scale $G_F^{-1}$ and
the unparticle propagates at the low scale $\mu_F$.
The fifth term is due to the difference of anomalous dimensions
of the charged unparticle operator in the effective Lagrangian \eqref{eq:leff}
and the neutral unparticle operator coupling to the Higgs VEV 
\eqref{eq:higgseff}, whether it acts as an enhancing or decreasing 
factor depends on the anomalous dimensions.
In a later paper \cite{BFRS} further dimensional analysis is explored. It is observed that when the coupling 
$\lambda$ multiplied by the supression scale $1/\Lambda_\cU^{\dU + (d_{\rm SM} -4)}$
is combined into a single scale $1/\Lambda_{(d_{\rm SM})}^{\dU + (d_{\rm SM} -4)}$
then under the assumption $1 < d_\dU < 2 < d_{UV}$, it is inferred that 
$\Lambda_2 < M_\cU < \dots < \Lambda_4 < \Lambda_3$
which seems counterintuitive at first sight since higher dimensional operators could receive an 
enhancement. 
 
Dimensional analysis is not very reliable. 
The construction of an explicit model would help to answer 
questions and presumably constrain the structure of
the effective Lagrangian \eqref{eq:leff}.

Possible
candidates are extensions of the standard model featuring near conformal dynamics, such as the walking technicolor theories. 
Those theories are close to an infrared fixpoint and hence have slow varying coupling constants.
A complete extension of the SM featuring walking dynamics
and its link to the underlying gauge theory has been given in 
Ref.~\cite{FFRS}.


\section{Critical discussion and conclusions}
\label{sec:con}

In this paper we have investigated the consequences
of the unparticle scenario in heavy flavour  physics. 
The new feature  is a CP odd or strong
phase that arises in the propagator as a consequence of 
the non integral scaling dimension. This gives rise to
very characteristic and novel CP violating phenomena.

The drawbacks of the scenario are that there is as yet no
concrete model and that it is not clear to what energies the 
scale invariant sector extends.
The lack of a model is overcome by parametrizing an effective 
Lagrangian, c.f. \eqref{eq:leff}, at the cost of many unknown
coefficients which have to be constrained. 
We have investigated the extension of the scale invariant sector
to lower energies
resorting to dimensional analysis. We have found that effects 
at the heavy flavour scales are possible provided the coupling of 
the unparticle field to the Higgs VEV is moderate at the scale $\Lambda_\cU$.
The effects are sizable for two reasons. 
Firstly the scaling dimension of the unparticle Lagrangian is more
relevant than the one of the weak Lagrangian and secondly  the effect of CP 
violation is linear and not shielded by a large SM background.

Bearing in mind the breaking of scale invariance we have
chosen decays where the unparticle propagates at a relatively
large scale.
The two examples  we have investigated are the decays 
$B^+  \to \tau^+ \nu$ and $B_d \to D^+D^-$.
In doing so we have assumed  the scale invariant sector  extends
to the scale $\sim 5\,{\rm GeV}$  for the former and to
$\sim 2\,{\rm GeV}$ for the latter.  We have not considered
decays into final state particles as in  for instance 
Ref.~\cite{Georgi1}
They would also lead to signals but we have assumed 
the unparticles to be weakly coupled.

We have chosen cases where the SM is described by a single
weak amplitude and the unparticles add a second weak amplitude
with strong phase allowing for the CP violation. In this sense our analysis does 
not differ from other model analyses with  two amplitudes.
The particularity of the unparticle scenario as compared
to other models is that it is an example where the
 large strong phase might be generated by the strength of
the coupling constant and that  the contribution to other 
(flavour)-channels is qualitatively different from other models, for example from those generating the strong phases through penguins.

The prediction of a CP asymmetry 
${\cal A}_{CP} = - C$ 
for leptonic decays seems a unique feature of 
the unparticle model, which has puzzling consequences to be 
discussed below.
The reparametrization invariant is the 
product of two quadratic invariants, one from 
the quark sector and one from the lepton sector.
As an example we have looked at  $B^+ \to \tau^+ \nu$
in conjunction with the constraints from the branching ratio.
Generic plots for the parameter space of the anomalous
dimension and the weak phase difference are shown
in Fig.~\ref{fig:btauBC}.
Maximal CP violation is possible for certain values of the 
parameter space. The current
experimental data is not yet strong enough to 
set absolute bounds.
Comments on flavour related 
decays are given in \ref{sec:rembtaunu}.
In particular the channel $D \to \mu \nu$ might be
of interest since  more events have been 
collected \cite{CLEO} than in $B \to \tau \nu$ \cite{BaBarBtaunu},
\cite{BelleBtaunu}. To the knowledge of the author there are no
experimental data available with bounds on CP asymmetries
in leptonic decays. Charge symmetry is usually implied in the analysis.

The investigation of the non-leptonic decay
$B_d \to D^+D^-$ was motivated by 
the large asymmetry $C_{D^+D^-}$ reported by
Belle  \cite{BelleDD}. 
We have neglected the penguin contribution
and treated the decay in naive factorization.
As compared to $B \to \tau \nu$ there is
a third observable, the time dependent
CP asymmetry $S_{D^+D^-}$. The latter agrees rather well with the 
SM predictions and sets constraints on $C_{D^+D^-}$.
It is possible though to find values where the CP
violation is maximal and satisfies the constraints 
of the branching ratio and the time dependent CP asymmetry.
As for $B \to \tau \nu$, plots for generic parameters are shown
in Fig.~\ref{fig:BDDCP}. It is encouraging that
for small ratios of effective couplings the constraints
from $S_{D^+D^-}$ allow for a large negative asymmetry $C_{D^+D^-}$
as reported by Belle whereas the opposite sign seems to 
be disfavoured. This fact is general to any analysis
with two amplitudes as outlined in section \ref{sec:genCP};
the unparticles just provide a scenario with two amplitudes and
possible large weak and strong phase differences. The true
meaning is that in the case where the decay is described by two
amplitudes, the sign of the Belle measurement is
more consistent than the opposite sign.
Discussions on U-spin and colour
related decays are given in section \ref{sec:disBDD}.
Let us emphasize two points from this section once more.
Generically we would expect a large asymmetry in 
$C_{D^+D^-}$ to be accompanied by a large asymmetry in 
the color related $C_{J/\Psi \pi_0}$. Currently the
experimental value 
$C_{J/\Psi \pi_0}^{\rm PDG} = -0.11(20)$ \cite{PDG}
is not conclusive and moreover $B_d \to J/\Psi \pi_0$ and on the 
theoretical side, complications arise due to non-factorizable contributions.
For $B_d$ decays the time dependent asymmetries are
typically proportional to $\sin(2\beta)$ or  $\sin(2\alpha)$,
the large angles of the $B_d$ triangle, and new physics contributions
are therefore hard to see. For $B_s$ decays, the mixing 
phase is $\phi_s \simeq 0$ and therefore the unparticle scenario
could give rise to sizable corrections. 
This would be particularly interesting for $B_s \to J/\Psi \phi$ which aims
at the extraction of the $B_s$ mixing phase $\phi_s$ at the LHCb. 

We have verified in section \ref{sec:cpt} 
that the novel CP-violation satisfies constraints from CPT-invarince,
namely the equality of the sum of partial rates, of the subclasses of final states
rescattering into each other, of particle and antiparticle.
Since the SM and no well-known new physics model predicts
a CP asymmetry for leptonic decays such as $B^+ \to \tau^+ \nu$
we have inferred that the compensating mode must be
due to unparticles.
As we have quantitatively verified, the compensating mode is 
$B^+ \to \cU^+$. This might appear surprising at first sight but is
possible since the unparticle does not have a definite mass 
but a continuous spectrum like a multiparticle state which
was one of the basic observations in Georgi's first paper
\cite{Georgi1}.

Clearly the unparticle scenario would benefit largely from the 
construction of an explicit model.
The question of the breaking of scale invariance and what a 
real\footnote{As opposed to  a virtual particle, on which we 
focused  throughout this paper.} 
unparticle in a laboratory experiment would mean
could be addressed and it would presumably also provide structural 
constraints on the coefficients of the effective Lagrangian.

\section*{Acknowledgments}

I am grateful to Ikaros Bigi, Oliver Brein, Luigi Del Debbio, Sakis Dedes, Stefan F\"orste, Uli Haisch, J\"org Jeckel, 
Francesco Sannino, Christopher Smith, Raymond Stora for discussions, 
to Nikolai Uraltsev for correspondence on the $B \to D$ form factor, to Paul Jackson, Sheldon Stone, 
Erika de Lucia and Roberto Versaci  for correspondence,
to Tom Underwood for help with figures
and to Lara Mary Turner for reading of the manuscript. 
Comments are welcome.

This work was supported in part by the EU networks
contract Nos.\ MRTN-CT-2006-035482, {\sc Flavianet}, and
MRTN-CT-2006-035505, {\sc Heptools}.

\appendix
\setcounter{equation}{0}
\renewcommand{\theequation}{A.\arabic{equation}}

\section{Some plots as a function of $\dU$}
\label{app:adu}

\begin{figure}[!h]
 \centerline{\includegraphics[width=2.5in]{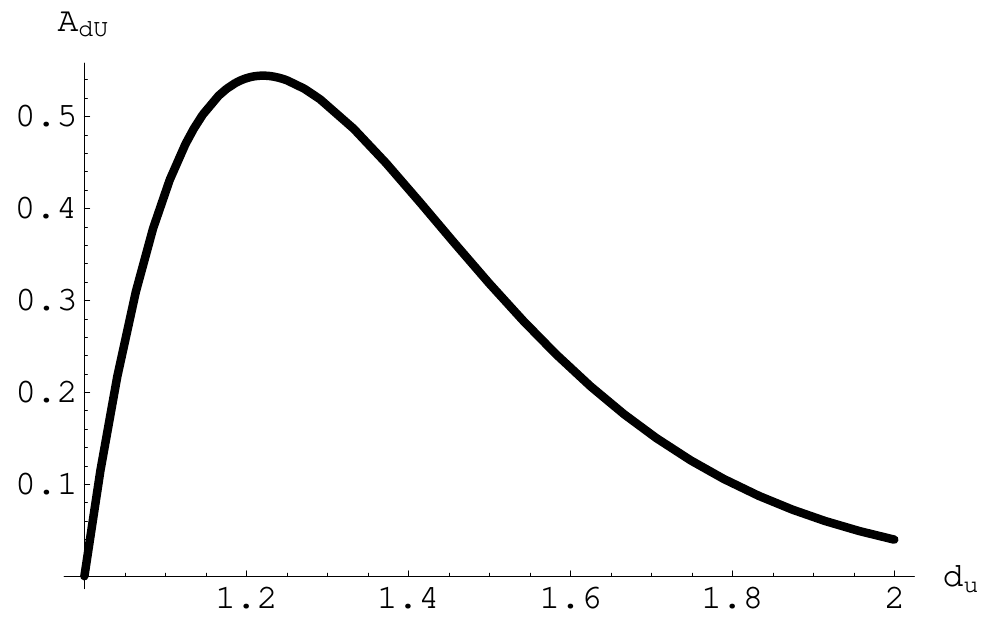}
 \includegraphics[width=2.5in]{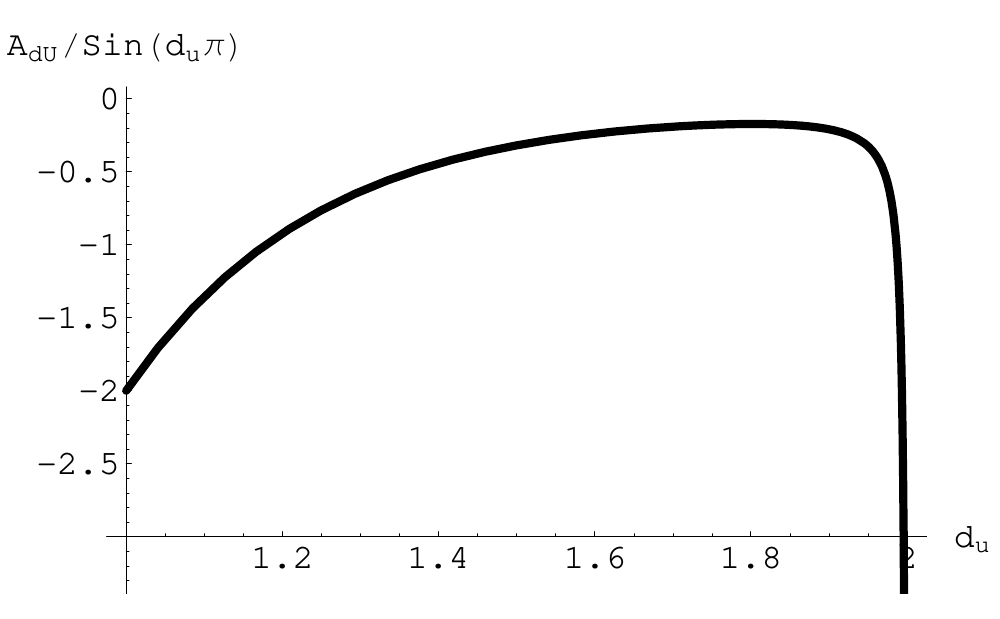}
 }
 \caption{\small  The phase space function $A_\dU$ \eqref{eq:adu} (left) and as appearing in the propagator  $A_\dU/\sin(\dU \pi)$ 
 \eqref{eq:prop} plotted against $\dU$ }
 \label{fig:adu}
 \end{figure}
 
 \begin{figure}[!h]
 \centerline{\includegraphics[width=2.5in]{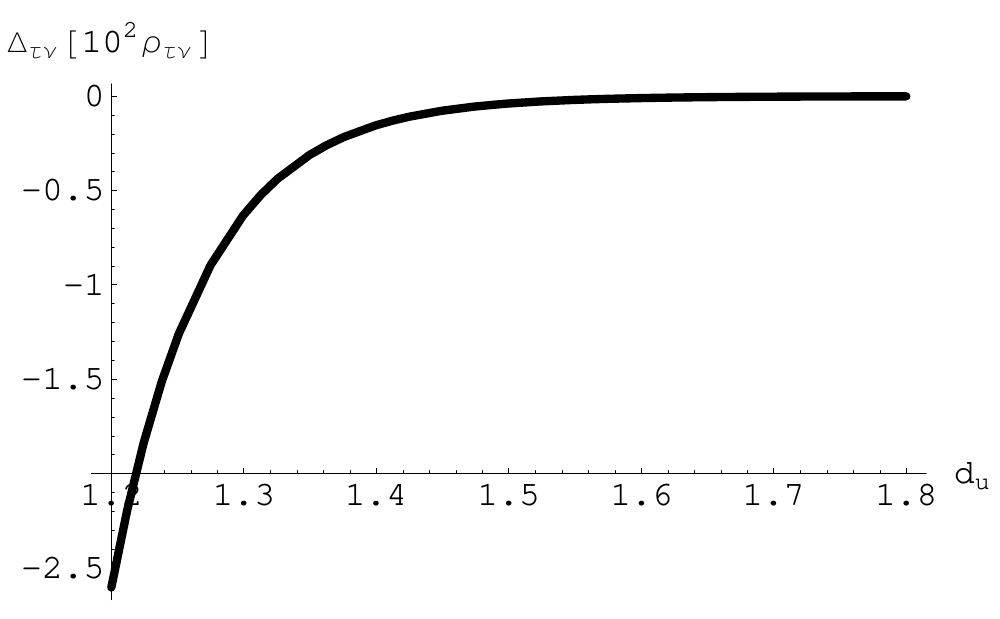}
 \includegraphics[width=2.5in]{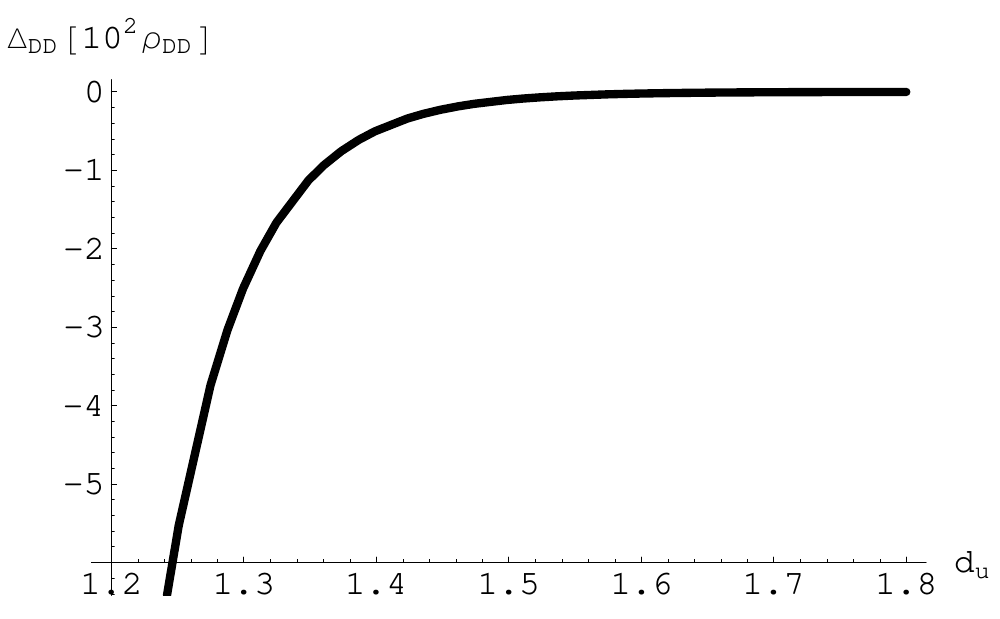}
 }
 \caption{\small The ratio of unparticle to SM model amplitude 
 $\Delta_{\tau\nu(\rm DD)}$ in units
 of the ratio of effective couplings $[\rho_{\tau\nu(\rm DD)} \cdot10^2]$  \eqref{eq:deltauexp}\eqref{eq:delDDex}, versus
 the anomalous dimension $\dU$. (left) $\Delta_{\tau \nu}$ 
 \eqref{eq:deltauexp}, (right) $\Delta_{\rm DD}$ \eqref{eq:delDDex}}
 \label{fig:deltas}
 \end{figure}

\section{Explcit results in coordinate space}

\subsection{The commuator}
\label{app:comm}

The commutator of the unparticle field
\begin{equation}
C(x) = \matel{0}{[O_\cU(x),O_\cU(0)]}{0}
\end{equation}
may be obtained from the time ordered product
\begin{equation}
C_F(x) =  \matel{0}{T O_\cU(x)O_\cU(0)}{0}
\end{equation}
via the general formula
\begin{equation}
C(x) =-2i{\rm sgn}(x_0){\rm Im}[C_F(x)] \quad.
\end{equation}
The correlation function $C_F(x)$ is obtained by Fourier transformation
of \eqref{eq:prop}, c.f. Ref.~\cite{gelfand}
\begin{equation}
\label{eq:feyn}
C_F(x) = \int \frac{d^4 P}{(2\pi)^4} e^{-iPx} (-i \Delta_\cU(P^2)) 
= - \frac{2^{\dU-4}}{\pi^2} \frac{A_\dU}{2 \sin(\dU \pi)} \frac{\Gamma(\dU)}{\Gamma(2-\dU)}(-x^2+i0)^{-\dU} \,.
\end{equation}
The imaginary part is
\begin{equation}
{\rm Im}[(-x^2+i0)^{-\dU}] = - \sin(\dU \pi)\Theta(x^2) (x^2)^{-\dU}
\end{equation}
and we obtain the commutator
\begin{eqnarray}
\label{eq:comm}
C(x) &=& -i {\rm sgn}(x_0)\Theta(x^2)  (x^2)^{-\dU} \sin(\dU \pi) \frac{2^{2 \dU-4}}{\pi^2} \frac{A_\dU}{ \sin(\dU \pi)} \frac{\Gamma(\dU)}{\Gamma(2-\dU)}
\nonumber \\[0.1cm]
&=& -i  {\rm sgn}(x_0) \Theta(x^2)  (x^2)^{-\dU}(\dU-1)   \frac{\Gamma(\dU+1/2)\pi^{1/2 - 2\dU}}{\Gamma(2\dU)\Gamma(2-\dU)}
\end{eqnarray}
The free field case $\dU \to 1$ 
\begin{equation}
\lim_{\dU \to 1+} C(x) =  \frac{-i} {2 \pi} {\rm sgn}(x_0) \delta(x^2)
\end{equation}
may be recovered by use of the formula $\lim_{\epsilon \to 0+} \epsilon |z|^{1-\epsilon} = \delta(z)$. Or for any
integer $n$
\begin{equation*}
 \lim_{\dU \to n+} C(x)  \sim  -i  {\rm sgn}(x_0) \delta^{(n-1)}(x^2) \,,
\end{equation*}
it is seen that the commutator has support on the light-cone only.

\subsection{The Thirring model - an  example with phase factor}
\label{app:thirring}
The Thirring model belongs to the class of exactly solvable
two dimensional models. It is a fermionic model with a vector current-current interaction. The exact solution of
the two point function was obtained by Johnson \cite{Johnson} as a
function of free fields
\begin{equation}
\matel{0}{T \Psi(x) \bar \Psi(0)}{0} = -i e^{- i 4\pi \gamma D_0(x)} G_0(x) \,,
\end{equation}
where 

\begin{equation}
D_0(x)  = \frac{-i}{4\pi} \log(-x^2+i0)  \qquad G_0(x) = \frac{1}{2\pi} \frac{\gamma_\mu{x^\mu}}{x^2-i0}
\end{equation}
are the free bosonic and fermionic Greens functions.
We have identified $\gamma = (\lambda^2/4 \pi^2)(1-\lambda^2/4\pi^2)^{-1}$, where $\lambda$ is 
the current-current coupling constant, as for instance
in  \cite{Wilson}. N.B. $\gamma > 0$ in accordance with \eqref{eq:d12} and $\dU = 1 + \gamma$.
We recover
\begin{equation}
\matel{0}{T \Psi(x) \bar \Psi(0)}{0} = \frac{i}{2\pi} \frac{\gamma_\mu{x^\mu}}{(-x^2+i0)^{1+\gamma}}
\end{equation}
the formula \eqref{eq:feyn}  in the fermionic case up to an overall normalization.
The phase factor arises due to resummation of thresholds at $x^2 > 0$.
Note that the overall normalization in a scale invariant theory is a matter of convention and
is hidden in the arbitrary scale factor in the logarithm of the free bosonic function 
$\log[(-x^2+i0)\mu^2]$.  In the notation of Eq.~\ref{eq:effUn}
the scale $\mu$ is proportional to the fixed point scale
$\Lambda_\cU$. This scales exhibits the phenomenon
of dimensional transmutation.


\section{Final state interaction in $B_d \to D^+D^-$  \\ consistent with naive factorization}
\label{app:iso}

In this appendix we shall obtain the isospin final state interaction phases 
within the naive factorization approach.
The isospin analysis  from $K \to \pi\pi$ and $B \to \pi\pi$, c.f. \cite{bigisanda} is transferable
to $\bar B_d \to D^+D^-$  \cite{Xing}. The $D$-mesons are $I = 1/2$ states. 
Angular momentum conservation implies that only $I=0$ and $I=1$ states
are formed as final states in the decay.
Denoting the amplitudes
\begin{equation*}
A^{+-} = {\cal A}(\bar B_d \to D^+D^-) \quad A^{+-} = {\cal A}(\bar B_d \to D^0D^0) \quad A^{+0} = {\cal A}(\bar B_u \to D^+D^0) \,,
\end{equation*}
isospin symmetry implies
\begin{equation}
A^{+0} = A_1 \quad A^{+-} = \frac{1}{2}(A_1+A_0) \quad A^{00} = \frac{1}{2}(A_1-A_0)
\end{equation}
from where the famous isospin triangle follows
\begin{equation}
A^{+0} = A^{+-} + A^{00} \quad .
\end{equation}
Let us introduce the following notation 
\begin{equation}
A_0 = Z_0 e^{i\delta_0} \qquad A_1 = Z_1 e^{i \delta_0} \quad ,
\end{equation}
where we have factorized the final state interaction phase
in the corresponding isospin channels. 

In the naive factorization approximation $Z_0 = Z_1 \equiv Z$ and
therefore
\begin{eqnarray}
\label{eq:isi}
A^{+0} &=&  Z \cos\Big( \frac{\delta_1-\delta_0}{2} \Big) e^{i (\delta_1-\delta_0)/2} \nonumber  \\
A^{00} &=&  i Z \sin\Big( \frac{\delta_1-\delta_0}{2} \Big) e^{i (\delta_1-\delta_0)/2} \nonumber  \\
A^{+0} &=&  Z e^{i \delta_1} \,.
\end{eqnarray}
The isospin triangle becomes  rectangular
\begin{equation}
|A^{+0}|^2 = |A^{+-}|^2 + |A^{00}|^2
\end{equation}
Two out of the three rates have been measured \cite{PDG}
\begin{equation}
\bar {\cal B}( B_d \to D^+ D^-)  = 1.9(6) \cdot 10^{-4} \qquad \bar {\cal B}( B^+ \to \bar D^0 D^+)  = 4.8(1) \cdot 10^{-4} \,.
\end{equation}
Neglecting irrelevant phase space effects we obtain from \eqref{eq:isi}
\begin{equation}
\label{eq:iso_final}
\cos^2 \Big( \frac{\delta_1-\delta_0}{2} \Big) \simeq 0.4(2) \,.
\end{equation}

\end{document}